\documentclass[sigconf,prologue, table]{acmart}

\usepackage{amsmath,amsfonts}
\usepackage[ruled,vlined,linesnumbered]{algorithm2e}
\usepackage{graphicx}
\usepackage{subfigure}
\usepackage{textcomp}
\usepackage{caption}
\usepackage{booktabs}
\usepackage{diagbox}
\usepackage{enumitem}
\usepackage{xcolor}
\usepackage{listings}
\newcommand{\scheduler}{Adaptive Resource-Moldable Scheduler (\texttt{ARMS})}
\newcommand{\sscheduler}{\texttt{ARMS}}
\newcommand{\rwss}{\texttt{RWS}}

\newcommand{\adws}{\texttt{ADWS}}
\newcommand{\xitao}{\texttt{Runtime\_X}}
\newcommand{\lags}{\texttt{ARMS-1}}
\newcommand{\armsn}{\texttt{ARMS-M}}

\newcommand{\intelskl}{\texttt{Intel Skylake}}

\usepackage{listings}
\usepackage{multirow}
\usepackage{efbox}
\efboxsetup{linecolor=blue,linewidth=1pt}
\usepackage{colortbl}
\usepackage{hhline}

\definecolor{codegreen}{rgb}{0,0.6,0}
\definecolor{codegray}{rgb}{0.5,0.5,0.5}
\definecolor{codepurple}{rgb}{0.58,0,0.82}
\definecolor{backcolour}{rgb}{0.95,0.95,0.92}

\lstdefinestyle{mystyle}{
    backgroundcolor=\color{backcolour},   
    commentstyle=\color{codegreen},
    keywordstyle=\color{magenta},
    numberstyle=\tiny\color{codegray},
    stringstyle=\color{codepurple},
    basicstyle=\ttfamily\footnotesize,
    breakatwhitespace=false,         
    breaklines=true,                 
    captionpos=b,                    
    keepspaces=true,                 
    numbers=left,                    
    numbersep=5pt,                  
    showspaces=false,                
    showstringspaces=false,
    showtabs=false,                  
    tabsize=2
}

\def\BibTeX{{\rm B\kern-.05em{\sc i\kern-.025em b}\kern-.08em
    T\kern-.1667em\lower.7ex\hbox{E}\kern-.125emX}}
\begin{document}
%
% \frontmatter
%
% \pagestyle{headings}
%
% \mainmatter

%candidate 1
% \title{Call to ARMS! Defeating communication overheads with an Adaptive Resource-Moldable Scheduler}

%candidate 2
%\title[Call to ARMS! The Adaptive Resource-Moldable Scheduler]{Call to ARMS! Combating inefficient task mapping decisions with an Adaptive Resource-Moldable Scheduler}
\title[ARMS - The Adaptive Resource-Moldable Scheduler]{Mitigating inefficient task mappings with an Adaptive Resource-Moldable Scheduler (ARMS)}
\author{Mustafa Abuljabbar}
\affiliation{%
   \institution{Chalmers University of Technology}
%   \streetaddress{30 Shuangqing Rd}
%   \city{Haidian Qu}
%   \state{Beijing Shi}
   \city{Gothenburg}
  \country{Sweden}
  }
\email{musabdu@chalmers.se}

\author{Mahmoud Eljammaly}
\affiliation{%
%   \institution{Gothenburg, Sweden}
%   \streetaddress{30 Shuangqing Rd}
   \city{Gothenburg}
%   \state{Beijing Shi}
   \country{Sweden}
  }
\email{jammaly@ieee.org}

\author{Miquel Peric\`as}
\affiliation{%
   \institution{Chalmers University of Technology}
%   \streetaddress{30 Shuangqing Rd}
%   \city{Haidian Qu}
%   \state{Beijing Shi}
   \city{Gothenburg}
   \country{Sweden}
  }
\email{miquelp@chalmers.se}

%candidate 3
% \title{Call to ARMS! Defeating locality/parallelism imbalance with an Adaptive Resource-Moldable Scheduler}

\begin{abstract}
Efficient runtime task scheduling on complex memory hierarchy becomes increasingly important as modern and future High-Performance Computing (HPC) systems are progressively composed of multi-socket and multi-chiplet nodes with nonuniform memory access latencies. Existing locality-aware scheduling schemes either require control of the data placement policy for memory-bound tasks or maximize locality for all classes of computations, resulting in a loss of potential performance. While such approaches are viable, an adaptive scheduling strategy is preferred to {enhance locality and resource sharing efficiency using a portable programming scheme}. In this paper, we propose the Adaptive Resource-Moldable Scheduler (\sscheduler\/) that dynamically maps a task at runtime to a %elastic 
partition {spanning one or more threads}% involved in the computation of the task
, based on the task and DAG requirements. 
% This is done with as little information as the software topology of input task (e.g. the Cartesian coordinates of the mesh points of the data). 
The scheduler builds an online {platform-independent} model for the local and non-local scheduling costs for each tuple consisting of task type (function) and task topology (task location within DAG). %depicted by its software topology address. 
We evaluate \sscheduler\/ using task-parallel versions of SparseLU, 2D Stencil, FMM and MatMul as examples. Compared to previous approaches, \sscheduler\/ achieves up to 3.5$\times$ performance gain over state-of-the-art locality-aware scheduling schemes. 

% we show that adopting a highly dynamic scheme outperforms locality maximizing schemes by up to $30\%$ for both classes of applications. Also, based on the schedule map of memory and compute-bound task types, \sscheduler\/ shows the ability to tune the resources oblivious of predefined user-level annotations.
\end{abstract}

\maketitle

\section{Introduction}
\label{sec:intro}
The trend towards larger many/multicore architectures adds pressure on the memory technology in order to sustain maximum performance. Such advancements typically target reducing latency, increasing and scaling data transfer bandwidth. 
% The transition from multicore to manycore processors exerts additional pressure on the memory resources. As a result, designers are integrating additional memory controllers on chips, resulting in non-uniform memory access latencies and bandwidths. To manage the corresponding decrease in yield, modern compute systems are increasingly built out of multiple chiplets, leading to a complex hierarchy of compute and memory elements. 
For a parallel program running on such platforms, performance is often subject to data access latency that varies based on where the data resides. This may range from local resources (e.g. L1/L2 caches) to uncore resources such as the L3 or DRAM. 
Furthermore, to increase the DRAM bandwidth and capacity, computing nodes increasingly leverage within-socket~\cite{amd-epyc2, fujitsu-a64fx2} and multi-socket~\cite{intel-scalable, amd-zeppelin} Non-Uniform Memory Access domains (NUMA) partitions, adding yet another level to the memory hierarchy. 
This makes the optimization of parallel programs rather more challenging as it is potentially impacted by nonuniform data access latency, memory channel bandwidth and resource contention. These challenges are often addressed by approaches that have the goal of maximizing data locality \cite{unat-tpds17}. 
Thus, advanced programming, runtime and compiler level techniques have been introduced to handle the resulting asymmetry in access speeds, to overcome penalties of non-local requests and maximize the utilized bandwidth~\cite{zoltan-icss-11, compiler-opt-locality-asplos-94, acar-locality, unat-tpds17}. 
% A parallel runtime system Task-based programs can be expressed in different ways, however, they are essentially realized by the parallel runtime system as a Direct Acyclic Graph, where nodes represent tasks and edges represent dependencies. 
% Task-based runtime systems offload the effort of scheduling (mapping tasks to cores/resources)
% Sub-NUMA Clustering~ \cite{knl,hpc_benchmarks_CLX_BDW}, and page-based NUCA (Non-Uniform Cache-Architecture) in multicore/manycore systems~\cite{hardavellas-isca09} among others. 
% As a result of adopting such technologies, the aggregate peak bandwidth essentially grows as a byproduct of multiple cores requesting independent memory channels. 
% From a core's perspective, it is well-established that data access latency depends on the location across the memory hierarchy (e.g. starting from register memory to local or remote DRAM). 
% Due to the impact of data locality on performance, advanced programming, runtime and compiler level techniques have been developed to handle the resulting asymmetry in access speed to reduce penalties of non-local requests and maximize utilized bandwidth~\cite{zoltan-icss-11, compiler-opt-locality-asplos-94, acar-locality, unat-tpds17}.
% Hence, investing in solutions such as the High Bandwidth Memory (HBM)~\cite{lee-hbm-isscc-14} or the Multi-Channel DRAM (MCDRAM)~\cite{sodani-knl-2015} becomes inevitable due to the large increase in memory requests issued by the processing units. %
% To maximize concurrency 

One of the expressive parallel programming paradigms are the task-based programming models~\cite{Nakashima2014_Massive_threads, Willhalm2008_TBB, wheeler-qthreads, duran-ppl11}. These are frequently used to express the parallelism of the program in the form of tasks.
The task scheduling problem (i.e. mapping tasks to execution units such as threads) is, however, nontrivial. This is particularly true with the growing complexity of task DAGs (Direct Acyclic Graph) and execution platforms. 
%that run on platforms involving many potential execution targets, which are dissimilar from a platform to another. 
Therefore, scheduling must be performed online by the task-based runtime schedulers. 

Many schedulers embraced by modern task-parallel runtime libraries   leverage random work-stealing as it achieves dynamic load-balance and increases the average core utilization \cite{blumofe-cilk}. Hence, work-stealing has become the de facto implementation choice of many task-parallel runtime systems. Nonetheless, random work-stealing assumes a flat view of the memory resources and is bound to suffer a considerable performance penalty, especially when scheduling tasks whose performance is influenced by data access latency and bandwidth.

For this reason, locality-aware extensions to work stealing have been proposed to mitigate the effects of non-local memory/cache access latency. These schedulers fall into three categories. In the first category, the task-DAG's input/output dependencies are analyzed to devise a locality-aware schedule~\cite{drebes-taco14, drebes-pact2016, Barrera-ics2018}. Tasks are then mapped to reduce a distance metric based on  a static description of the hardware topology, which explains how the different hardware components (such as caches, memory controllers, cores, threads, etc.) are organized. A second category focuses on creating work-stealing regions with low latency memory access. This is achieved by leveraging locality hints from the programmer to aggregate the tasks into execution places (i.e.~collections of cores) that can be configured to match the hardware architecture such as threads, cores or sockets. Examples of this approach include LAWS~\cite{guo-ipdps10}, ADWS~\cite{ShiinaSC19_ADWS}, SWAS~\cite{swas-tzilis-icppw-17} and Olivier et al.~\cite{Olivier_IJHPCA12_OpenMP}. In the third category, the task data allocation is controlled by the runtime to  evenly distribute the data over the available NUMA domains, then schedule tasks close to their known data regions~\cite{chen-ics14}. 

While locality-driven execution is a key factor to reducing data access latency and increasing performance, the overall application performance gain is subject to %complex 
elements that are not clearly addressed by such schemes. 
Consider, for example, the arithmetic intensity (AI, i.e. flops per byte ratio). Application kernels with high AI are less impacted by data placement and can benefit from maximizing the compute throughput of all available execution units. This can be achieved by using simple greedy schemes such as work stealing. 
Hence, such applications do not benefit from the scheduling schemes that start by 
%However, the latter appears to be orthogonal to the counterpart that starts 
by maximizing locality and then applying 
%handles the preventable work time inflation by 
work stealing~\cite{ShiinaSC19_ADWS}, or schemes that limit their scope to tasks that have higher LLC miss rates~\cite{drebes-pact2016}.
Another key element to achieving higher application performance is maximizing the task data reuse and reducing resource requirements (e.g. bandwidth). 
This is specifically important for fine-grain task whose data size is within the lower level caches. Finally, it is not clear how such schedulers behave in events of lower DAG parallelism, which can be a result of unwinding the recursion in common divide-and-conquer DAGs that are a special case of fully strict DAGs~\cite{blumofe-cilk}. 
% can benefit from maximizing cache locality.
In light of these elements, a scheduler that dynamically adapts to the task and DAG requirements (e.g. flops, channel bandwidth, cache reuse, parallelism, criticality) can obtain high performance on a variety of platforms while at the same time simplifying the programmer's task. 
% This is because it does not need to be equipped with a predefined knowledge of the workload such as computational or locality hints, and does not make any assumption on task resource requirements. Existing schemes require computational hints or locality hints, which are not readily available and/or require an understanding of the runtime internals. Also, to the best of our knowledge, there is no solution that addresses the known tradeoff between locality and parallelism at runtime to achieve higher application performance. %We believe that there is a need for 
% A scheme that dynamically aggregates resources to a task (i.e. exploiting internal task parallelism) based on its locality and resource requirements, and increase inter-task parallelism (i.e. using work-stealing) for tasks that are not locality sensitive would greatly contribute to addressing the locality problem for current and future NUMA-based architectures. 

This paper proposes \sscheduler\/, the Adaptive Resource-Moldable Scheduler. \sscheduler\/ dynamically builds a model of the performance of each application task on the system resource partitions, such as the cores that share cache levels, NUMA nodes, sockets, etc. {The model used, herein, serves as a predictor for the performance of a task on the available system resource partitions, and guides the scheduling decisions.} 
It is created at runtime for each task based on its location in the task DAG topology.
%mapped by its \textit{Software Topology Address} (STA). The STA 
{Software} topology information is used as a portable key by the runtime to map the logical location of task's input (e.g.~the Cartesian coordinates or the matrix indices) to a physical core. In the absence of topology information, the scheduler automatically assigns a relative address to the task based on its location in the DAG (depth and breadth). 
Hence, \sscheduler\/ is able to model the effects of scheduling a task on multicore partitions within or outside of its data region (e.g., NUMA node), identified by the task's specific topology location. Then, it uses this knowledge to create a schedule that reduces the per-task parallel cost.
In addition, \sscheduler\/ supports not only traditional 1:1 mapping (i.e. a single task to a single thread) but also 1:M mapping (i.e. moldability when 1 task is assigned M resources via worksharing). {A moldable task  encompasses an internal scheduler that assigns the work partitions to threads within the task function. This, in essence, represents a Single-Program-Multiple-Data (SPMD) region.} % by leveraging moldability. 
%Thus, each task has an internal parallelism. 
The flexibility in the mapping allows, for example, a memory-bound task to leverage a higher memory bandwidth by using more resources (threads), while a resource intensive task can set the number of threads to match its cache requirements. Also,  {the number of threads} assigned to a task is dynamically changed based on the DAG's parallelism.
\noindent

The main contributions of this work are as follows:
\begin{itemize}
    \item We identify the key parameters that constitute a locality adaptive performance model for a task, which are the task work function and the topology information.
%    This allows to create a high-performance schedule that adapts to the task and application requirements.
    \item We map the parameters to a model that captures the performance on the available execution places, which is used by \sscheduler\/ via a novel resource selection algorithm to deliver schedules that achieve good balance between locality and parallelism.
    %to know the most efficient places to map the task to. 
    \item We conduct a thorough analysis of the effectiveness of \sscheduler\/ against the state-of-the-art schedulers. This study reveals that even where there is a high DAG concurrency, resource aggregation (i.e. with 1:M mapping) is inevitable for memory intensive tasks, and in events of changing DAG parallelism. It also demonstrates that for tasks with high-compute intensity, locality maximizing schedules should be avoided. 
\end{itemize}

% \begin{itemize}
%     \item 
%     %A novel scheme to map the task's work function and data location to a performance model on the underlying system's execution places. 
%     A novel scheme to map the task’s work function and data location to the underlying system’s execution places by modeling the performance of scheduling decisions.
%     This allows to create a high-performance schedule based on the task and application requirements.
%     \item A thorough analysis of the effectiveness of \sscheduler\/ compared to locality-aware and traditional scheduling schemes across different tasks types, workload sizes and DAG parallelisms. 
% \end{itemize}
% \begin{itemize}
%     % \item The \scheduler: an adaptive scheduler that discovers when to localize or globalize the scheduling decisions, and adopts an online history-based model developed for each pair of STA and task. 
%     \item A novel approach to dynamically tune how much locality to preserve for each task %    the degree of locality preserving for a task 
%     (in terms of where to execute each task) % execution place and size) 
%     by employing an online performance model paired with architecture-independent topology information. 
%     \item A thorough analysis of the applicability of locality-aware and traditional scheduling techniques for representative classes of DAGs, which include iterative and recursive task graphs.% including X, y and z
%     %\item We analyze the impact of task parallelism and data locality on the decisions by the ARMS scheduler. The study shows that....
% \end{itemize}

The rest of the paper is organized as follows: Section~\ref{sec:background} introduces the scope and a few useful scheduling terminology used in this paper. Section~\ref{sec:lagres} describes the scheduling algorithm and the components used to develop the dynamic locality scheduling decisions.
% Section~\ref{sec:implementation} highlights the implementation challenges, followed by 
In Section~\ref{sec:methodology}, we describe the evaluated applications and the experimental methodology. Finally, Section~\ref{sec:eval} evaluates the scheduler with respect to locality-aware and traditional techniques. Section~\ref{sec:related_work} highlights the related work in the field of locality-aware work-stealing runtimes. Finally, Section~\ref{sec:conclusion} concludes the paper. 
% \begin{minipage}[t]{0.75\columnwidth}
% \begin{figure}[H]
% \begin{lstlisting}[language=C++,caption={Pseudocode for a 1D N-Body task.},captionpos=b, label={listing:motivation}, basicstyle=\scriptsize,frame=single]
% void n_body(pos_target[out], pos_source[in], 
%             time_step, size) { 
%   // partition over threads to set start 
%   // and end to support the moldable case
%   for(i = start; i < end; ++i) { 
%     Fx = 0
%     for(j = 0; j < size; ++j) {
%       dx      = pos_target[i] - pos_source[j];
%       invDist = rsqrt(dx * dx + SOFTENING)
%       Fx += dx * invDist
%     }
%     pos_target[i] += // use Fx and timestep
%   }
% }

% \end{lstlisting}
% \end{figure}
% \end{minipage}
%  \hspace{0.2cm}
% \begin{minipage}[t]{0.25\columnwidth}
% \vspace{-0.5cm}
% \begin{figure}[H]
%      \centering
%     \includegraphics[scale=0.27]{figs/motivational/motiv_dag_1.pdf}
%   \caption{\newline Motivational N-Body DAG.}
%     \label{fig:motiv_dag}
% \end{figure}
% \end{minipage}
% \efbox{
\begin{figure}[t]

\begin{lstlisting}[language=C++,caption={1D N-Body task (for 1-unit mass bodies).},captionpos=t, label={listing:motivation}, basicstyle=\scriptsize,frame=single, framerule=0.5pt]
void n_body(pos_target[in/out], pos_source[in], time_step, size) { 
  // set partition start and end 
  [start, end] = get_partition(_tid, _resource_width, size);
  for(i = start; i < end; ++i) { 
    Fx = 0
    for(j = 0; j < size; ++j) {
      dx = pos_target[i] - pos_source[j];
      invDist = rsqrt(dx * dx + SOFTENING)
      Fx += dx * invDist
    }
    pos_target[i] += // use Fx and timestep to update position
  }
}

\end{lstlisting}
\end{figure}
% }
\begin{figure}[t]
     \centering
   \efbox{\includegraphics[scale=0.3]{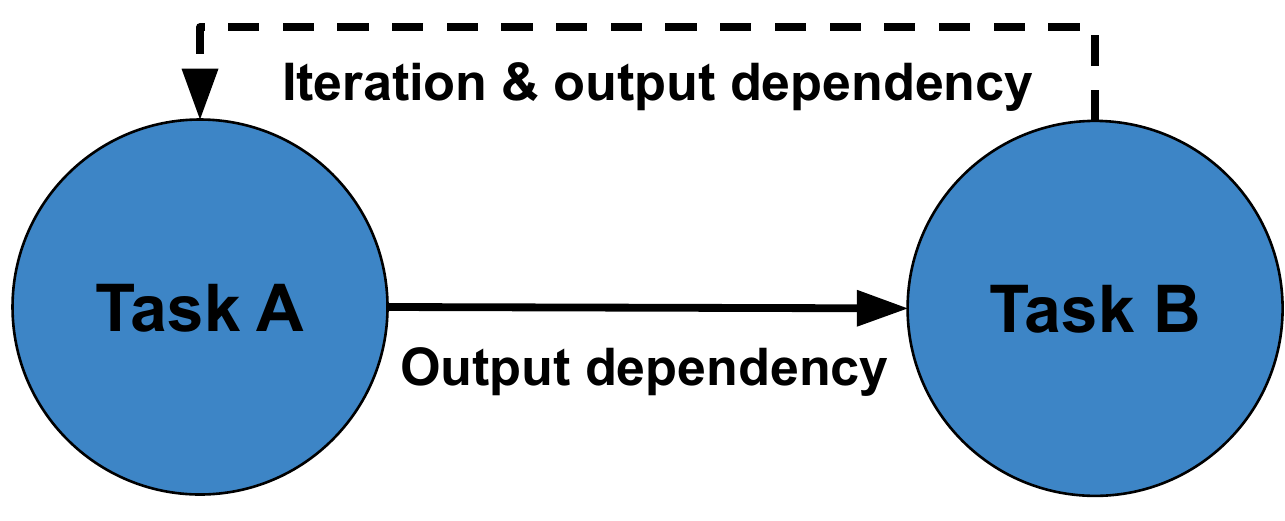}}
   \caption{Motivational N-Body DAG.}
    \label{fig:motiv_dag}
\end{figure}

% \begin{figure}[t]
%     \centering
%     \includegraphics[scale=0.5]{figs/motivational/motiv_dag.pdf}
%     \caption{Motivational DAG.}
%     \label{fig:motiv_dag}
% \end{figure}
% \begin{figure}[t]
% \centering
% \subfigure[Intra-NUMA access scenario]{ \includegraphics[width=0.47\columnwidth]{figs/motivational/local-case.pdf}
% \label{fig:nbody_size_vs_flops_no_mold}} 
% \subfigure[Cross-NUMA access scenario]{ \includegraphics[width=0.47\columnwidth]{figs/motivational/remote-case.pdf}
% \label{fig:nbody_size_vs_flops_mold}} 
% \caption{The executing thread.}
% \label{fig:nbody_size_vs_mflops_work_time}
% \end{figure}
% \usepackage{graphicx}
% \usepackage{colortbl}
% \usepackage{multirow}
% \usepackage{hhline}

\subsection{Motivation}
\label{subsec:motivation}

\begin{figure}[t]
\centering

\subfigure[Core MFLOP/s for a direct  N-Body DAG chain when tasks are  \textbf{not molded} {(i.e. single-threaded)}]{
%\efbox{
\includegraphics[width=0.42\columnwidth]{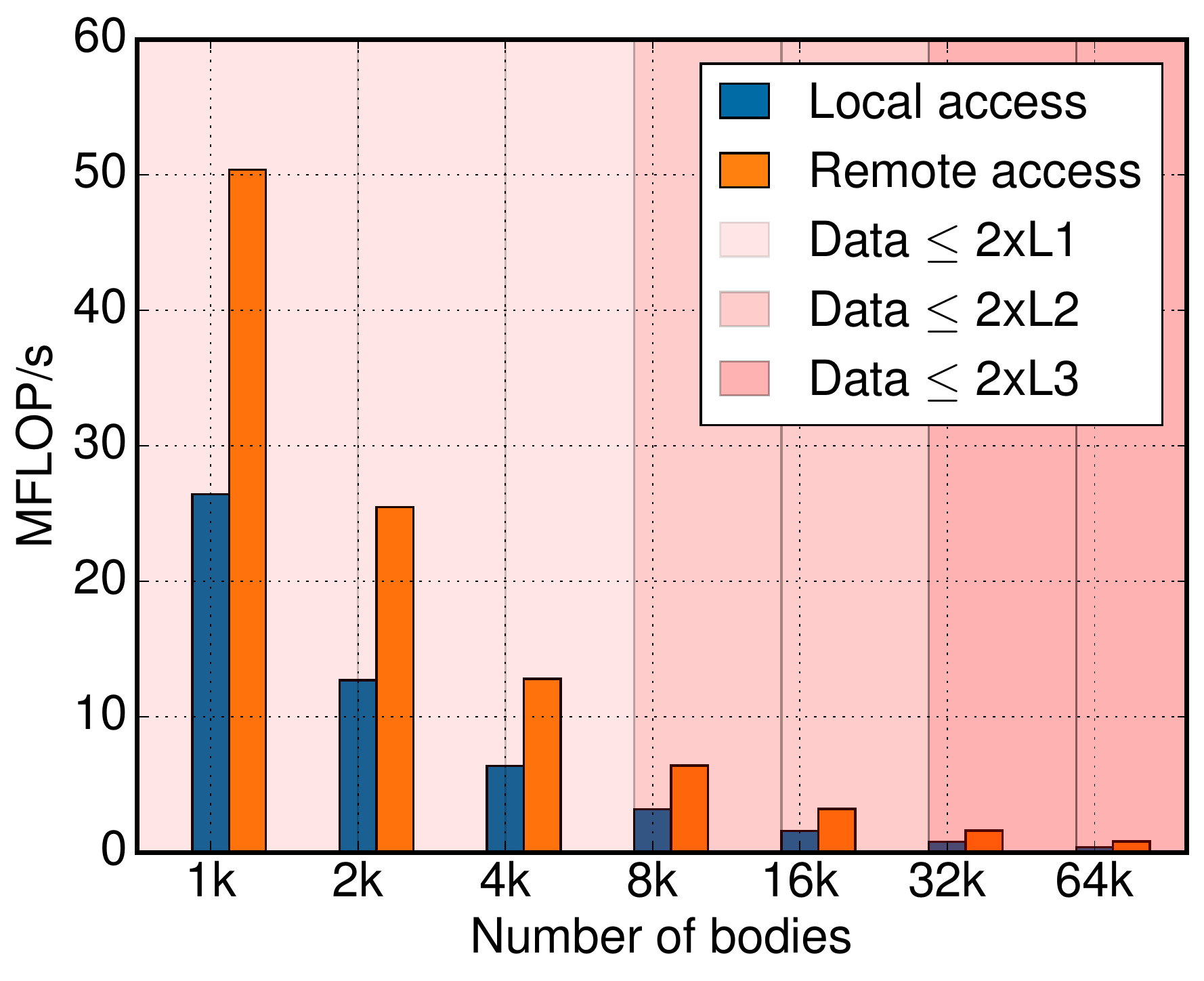}
%}
\label{fig:nbody_size_vs_flops_no_mold}}\hspace{3pt} 
\subfigure[Core MFLOP/s for a direct N-Body DAG chain when tasks are \textbf{molded} {to two threads}]{ 
% \efbox{
\includegraphics[width=0.42\columnwidth]{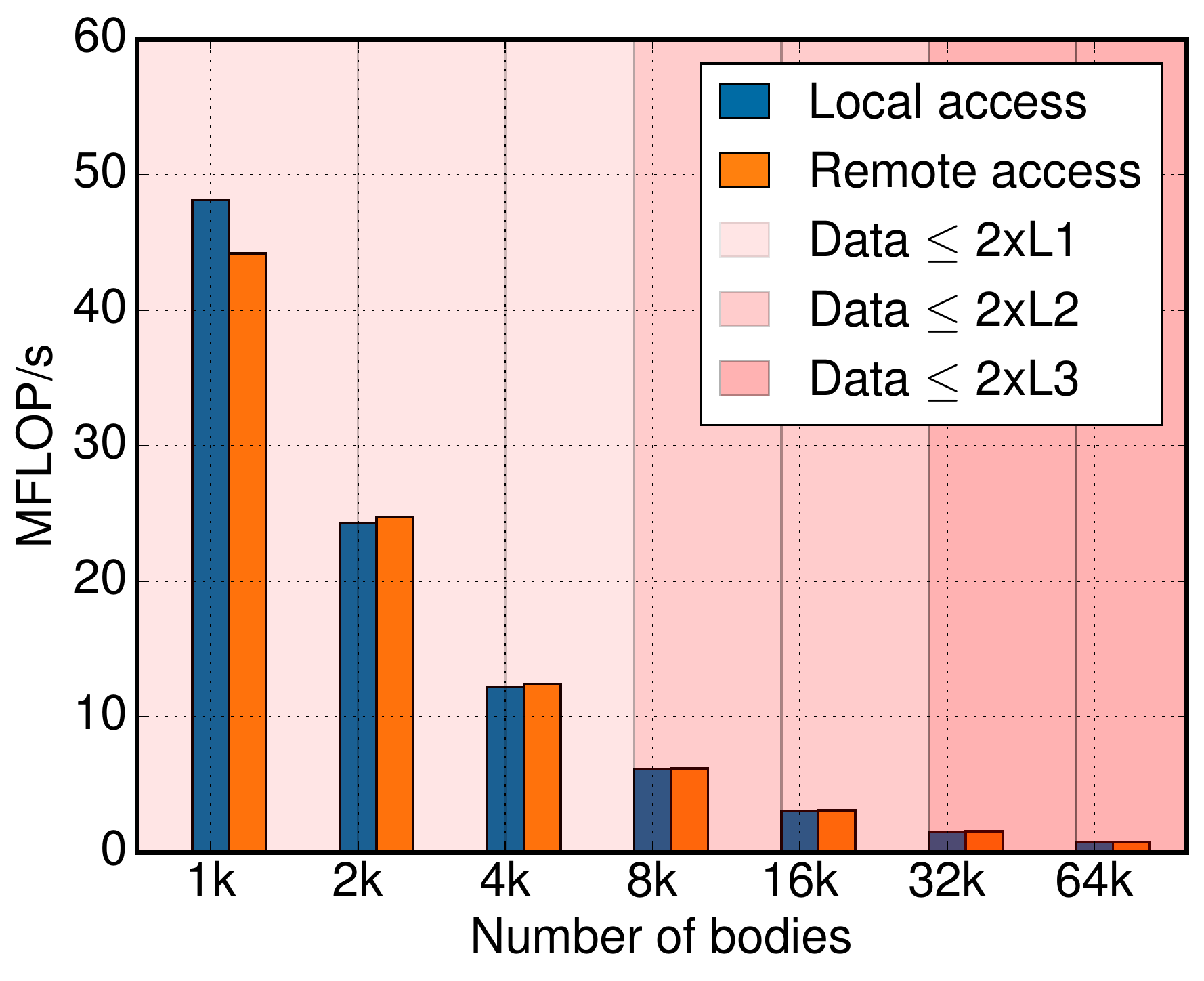}
%}
\label{fig:nbody_size_vs_flops_mold}}

% \\
% \subfigure[Cumulative core work time for a direct N-Body DAG chain when task is not molded to two cores]{ \includegraphics[width=0.47\columnwidth]{figs/motivational/intel-skl/test_time_compute_no_mold_time.pdf}
% \label{fig:nbody_size_vs_work_time_no_mold}} 
% \subfigure[Cumulative core work time for a direct N-Body DAG chain when task is molded to two cores]{ \includegraphics[width=0.47\columnwidth]{figs/motivational/intel-skl/test_time_compute_mold_time.pdf}
% \label{fig:nbody_size_vs_work_time_mold}} 
\caption{Discussing the dynamic factors that impact the core performance which include task size, data location, and task molding (averaging 30 runs) on a dual-socket Intel Skylake.}
\label{fig:nbody_size_vs_mflops_work_time}
\end{figure}

Figure~\ref{fig:nbody_size_vs_mflops_work_time} shows the impact of data locality %(considering NUMA) 
and task work size on the core performance using a compute-intensive {chain of N-Body tasks} running on a dual-socket 16-core Intel Skylake (described in Section~\ref{sec:methodology}) {. The study compares single-threaded (non-moldable) to dual-threaded tasks (a molding option). The experimental DAG chain is shown by Figure~\ref{fig:motiv_dag}, where the dotted line represents both an iteration and an output dependency from Task B back to Task A. Caches are flushed after each iteration to analyze the effect of streaming data across local and remote NUMA domains. The cache reuse happens within the iteration. Then, the chain is terminated after 1000 iterations to get a sustainable flops rate}. Performance is expressed using the core flops (MFLOP/s). The average is taken across multiple runs to give a statistical guarantee over reproduciblity. The shaded regions {in Figure~\ref{fig:nbody_size_vs_mflops_work_time}} depict the different levels of the caches where the total working set fits.
Listing~\ref{listing:motivation} briefly highlights the benchmark evaluated herein. It shows a direct N-Body ($N^2$) computations using 2 threads. {The output of \texttt{task A} is input to \texttt{task B} and so on. The \texttt{start} and \texttt{end} indices mark the partition that the task works on in case of moldable execution. The task's data} pointer {(i.e. \texttt{pos\_target})} is either pinned to the local NUMA or remote NUMA. {Specification of computation and data locations are shown in Table~\ref{tbl:motiv_table}. The ``not molded'' column shows the configurations used for Figure~\ref{fig:nbody_size_vs_flops_no_mold}, whereas the ``molded'' column shows the configurations used for Figure~\ref{fig:nbody_size_vs_flops_mold}. So for the chain of executions of \texttt{task A} followed by \texttt{task B}, we specify the computation thread id, and the NUMA id of the cell data (the N-Body domains are usually structured as trees). In this case, the cell data is \lstinline{pos_target} as per Listing~\ref{listing:motivation}. For example, in the ``Remote access'' scenario, the \lstinline{pos_target} output of \texttt{task B} located on NUMA node 1 would be \lstinline{pos_source} input to \texttt{task A} that executes on a thread 0 residing on remote NUMA node 0.
}
\begin{table}[t]
\caption{Task computation and data locations for motivational DAG (Figure~\ref{fig:motiv_dag}) discussed by Figure~\ref{fig:nbody_size_vs_mflops_work_time}. The table shows the configurations for the moldable and non-moldable cases for the local and remote scenarios.}
\centering
\arrayrulecolor{black}
% \efbox{
\resizebox{\linewidth}{!}{%
\begin{tabular}{|c|c|c|c|c|c|} 
\hline
\rowcolor[rgb]{0.753,0.753,0.753} Scenario & Mode & \multicolumn{2}{c|}{\textit{Not molded}} & \multicolumn{2}{c|}{\textit{Molded}} \\ 
\hline
{\cellcolor[rgb]{0.753,0.753,0.753}} & \multicolumn{1}{l|}{{\cellcolor[rgb]{0.753,0.753,0.753}}Task} & \multicolumn{1}{l|}{\textit{Task A }} & \multicolumn{1}{l|}{\textit{Task B }} & \multicolumn{1}{l|}{\textit{Task A }} & \multicolumn{1}{l|}{\textit{Task B }} \\ 
\hhline{|>{\arrayrulecolor[rgb]{0.753,0.753,0.753}}->{\arrayrulecolor{black}}-----|}
{\cellcolor[rgb]{0.753,0.753,0.753}} & \multicolumn{1}{l|}{{\cellcolor[rgb]{0.753,0.753,0.753}}Execution thread id(s)} & 0 & 0 & 0, 1 & 0,1 \\ 
\hhline{|>{\arrayrulecolor[rgb]{0.753,0.753,0.753}}->{\arrayrulecolor{black}}-----|}
\multirow{-3}{*}{{\cellcolor[rgb]{0.753,0.753,0.753}}\textbf{Local access}} & {\cellcolor[rgb]{0.753,0.753,0.753}}NUMA node id for data & 0 & 0 & 0 & 0 \\ 
\hline
{\cellcolor[rgb]{0.753,0.753,0.753}} & \multicolumn{1}{l|}{{\cellcolor[rgb]{0.753,0.753,0.753}}Execution thread id(s)} & 0 & 16 & 0,1 & 16,17 \\ 
\hhline{|>{\arrayrulecolor[rgb]{0.753,0.753,0.753}}->{\arrayrulecolor{black}}-----|}
\multirow{-2}{*}{{\cellcolor[rgb]{0.753,0.753,0.753}}\textbf{Remote access}} & {\cellcolor[rgb]{0.753,0.753,0.753}}NUMA node id for data & 0 & 1 & 0 & 1 \\
\hline
\end{tabular}
}
% }
\label{tbl:motiv_table}
\end{table}

% to the worker threads (i.e. \texttt{thread0} and \texttt{thread1}). 
We make two observations here: First, if we consider {the ``molded'' case shown by} Figure~\ref{fig:nbody_size_vs_flops_mold}, NUMA-aware computation ({\texttt{i.e. Local access}}) for this benchmark is beneficial only for the finest grain case (input size = \texttt{1k}). Second, when the task is not molded (as shown in Figure~\ref{fig:nbody_size_vs_flops_no_mold}), there is no gain in preserving NUMA locality. Interestingly, in the absence of moldability, the most locality preserving approach{, that is running dependent tasks on thread 0 and allocating on the local NUMA node,} does not perform well on average, as shown by the blue bars in Figure~\ref{fig:nbody_size_vs_flops_no_mold}. This is the case when computation is done next to the data in physical memory using a single thread to maximize L1 cache reuse. So, in practice such a choice incurs an evident performance degradation in certain cases. {In this case, ``remote'' wins because of the interleaved access to multiple memory channels (0,1) from tasks A and B when data is distributed over 2 memory domains.}
As these observations vary based on the application, they make dynamic task scheduling decisions even more desirable, since the best mapping depends on the underlying platform, class of computation, and granularity of task among other attributes. 
The main hypothesis behind ARMS is that the impact of scheduling decisions can be estimated via an online performance model that is aware of the task type and of topology information that encodes the data location. 

\begin{figure}[t]
    \centering
    \includegraphics[scale=0.5]{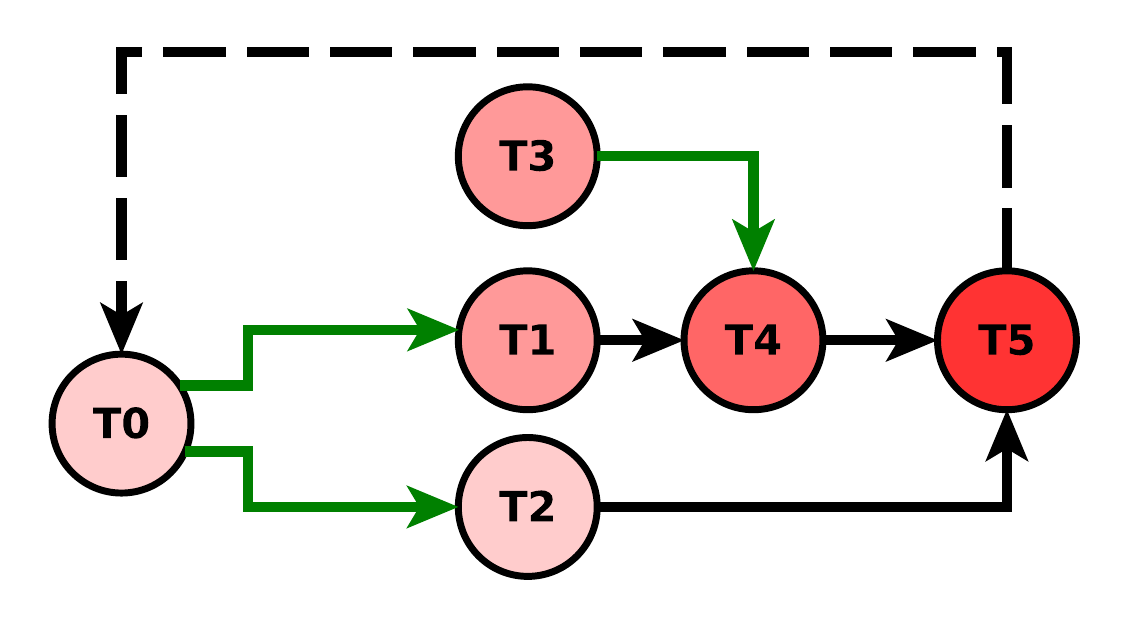}
    \caption{Sample generic DAG discussed in this paper. Green line indicates execution dependencies. Black line indicates data dependencies. Dotted line is an iteration (not a cycle).}
    \label{fig:dag}
\end{figure}
\section{Scope and Background}
\label{sec:background}
\begin{figure}
    \centering
    \includegraphics[scale=0.5]{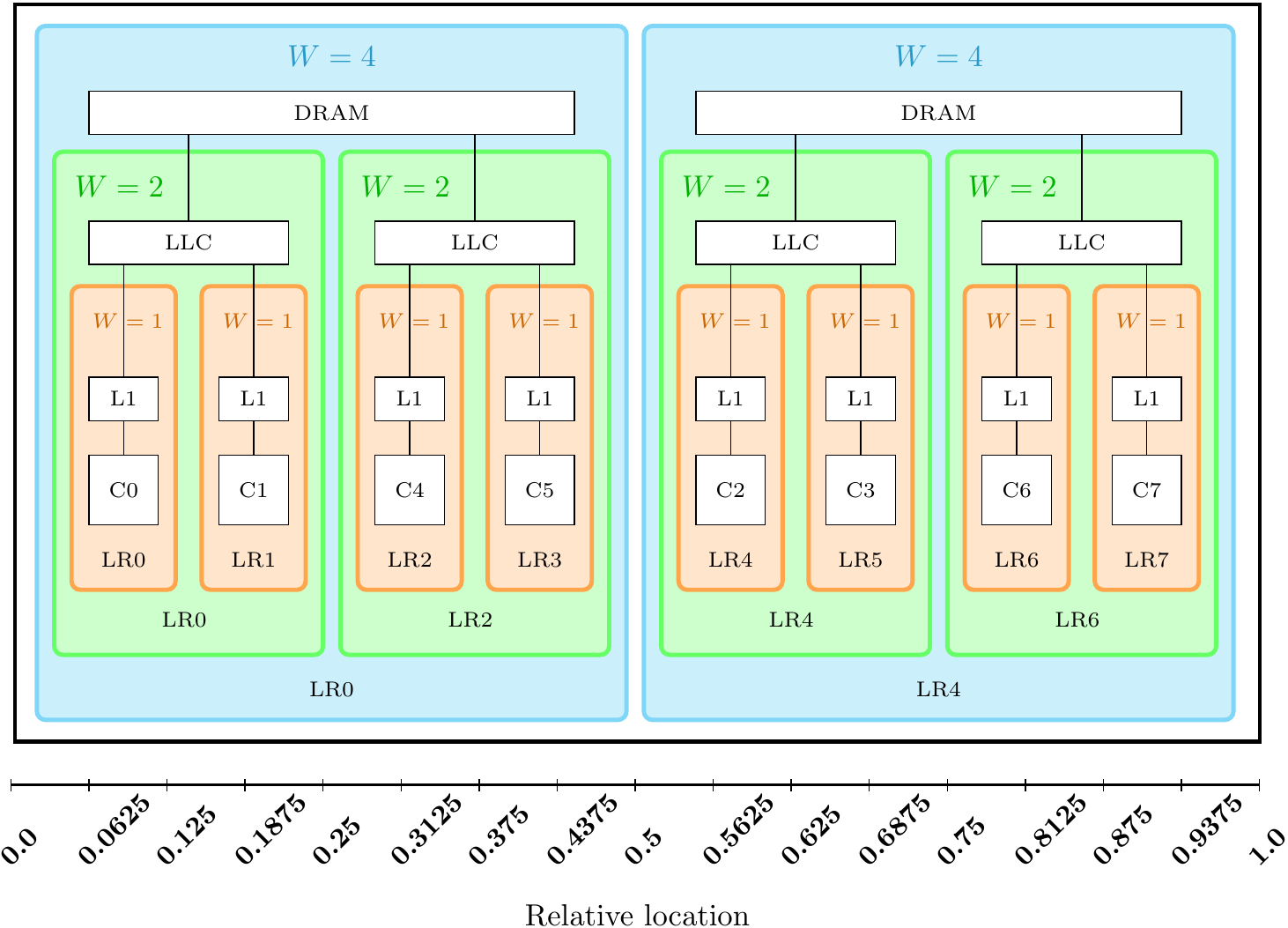}
    \caption{An example mapping of relative topological locations. {For instance, 0.125 maps to an initial worker id of 1 (0.125$\times$8). This worker falls in execution places: $R_0=[LR=0,W=1]$, $R_1=[LR=0,W=2]$ and $R_2=[LR=0,W=4]$}.}
    \label{fig:resource_partitions}
\end{figure}

This work targets a general form of DAGs with iterative or recursive structure, where nodes represent tasks and edges are either execution or data dependencies as shown by the synthetic example in Figure~\ref{fig:dag}. {The black edges} refer to direct data dependencies (i.e. there is an edge from $T_a$ to $T_b$ $\iff$ $T_b$ directly reuses output data from $T_a$). {The green edges} refer to execution dependencies (i.e. there is an edge from $T_a$ to $T_b$ $\iff$ $T_b$ cannot start unless $T_a$ has completed execution). For example, a task-based version of post-order recursive traversal would not visit a node unless all descendants have been traversed. The dotted line indicates an iteration, which is a concatenation {of the DAG to itself for an $X$ number of iterations}. This edge does not indicate a cycle {as the execution terminates after the program iterations have elapsed}. The gradient change in red color indicates a different \texttt{task type}. {A \texttt{task type} in the context of this work represents the task's work function}. Figure~\ref{fig:resource_partitions} depicts the possible \texttt{Resource Partitions} ($R_p$) denoted by the \texttt{Leader} ($LR$) logical thread id, which is the thread with the smallest id in the partition, and the \texttt{Width} ($W$) of the partition, which is the number of workers {(i.e. logical threads)} involved. $T_a$ can be mapped to a work-sharing region with $W>= 1$ $\iff$ $T_a$ is moldable.
% A multi-threaded task may benefit from moldability or even a single-threaded task that performs better with exclusive access to shared resources (e.g. caches). 

\section{\scheduler}
\label{sec:lagres}
In a parallel system, if the available resource partitions are denoted by $R_p=\{R_0,R_1,...,R_{p-1}\}: R_i=[LR,W]$, then each invocation of \sscheduler\/ (denoted by $S$) on a task $\tau$ should return $R_x$ ($S(\tau)=R_x \in R_p$). Therefore, this section describes the different components of \sscheduler\/. In Section~\ref{sub:sta}, we define the concept of STA (Software Topology Address) and show how to construct STAs in order to differentiate between the performance models based on locality. Then in Section~\ref{sub:mold}, we show how we build the moldable resource partitions that are to be analyzed by the model. Section~\ref{sub:online_perf} highlights the resource selection algorithm. These components are necessary to determine what constitutes an efficient resource partition to schedule a specific task based on an online performance model.
% ($\{(\tau,S(\tau)):\tau \in T, S(\tau) \in R_p\}$).
\begin{figure*}[t]
\centering
\subfigure[Sample Cartesian layout]{ \includegraphics[width=0.2\textwidth]{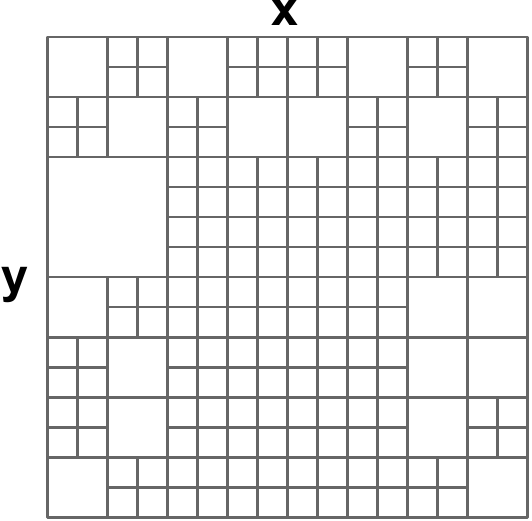}
\label{fig_cart_layout}} \hspace{2pt}
\subfigure[Coarsened linear mapping of task data coordinates (e.g., Morton order)]{ \includegraphics[width=0.185\textwidth]{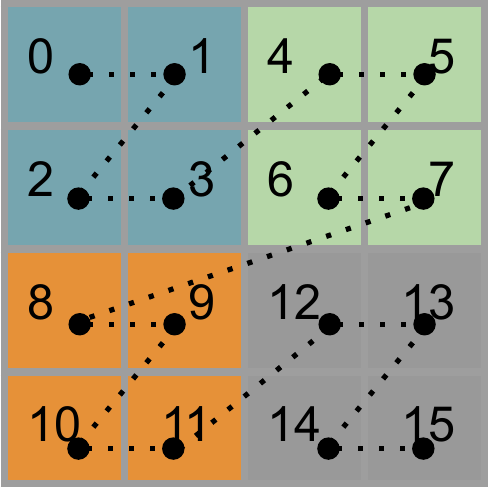}
\label{fig_linear_mapping}}  \hspace{2pt}
\subfigure[Creating performance model per locality (linear map)]{ \includegraphics[width=0.2\textwidth]{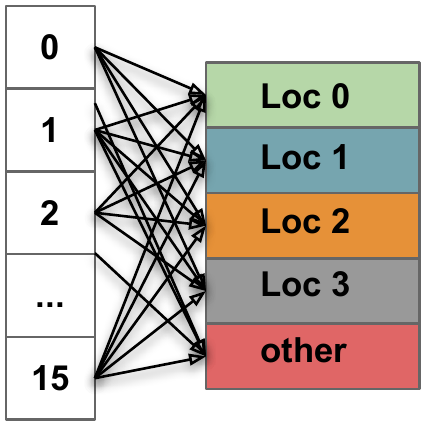}
\label{fig_perf_model}} \hspace{2pt}
\subfigure[Mapping tasks based on performance model]{ \includegraphics[width=0.2\textwidth]{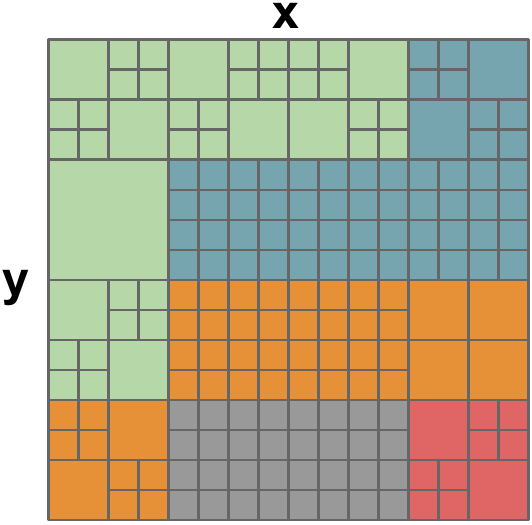}
\label{fig_task_mapping}} 
\caption{Overview of \sscheduler}
\label{fig:overview}
\end{figure*}
\subsection{The Software Topology Address Construction}
\label{sub:sta}
Among the common techniques in constructing logical representations of spatial domains is using graph, geometric or algebraic forms. For example, continuous air particles, once discretized, can be expressed as mesh topology points accessed by their Cartesian coordinates. This matrix-free representation is useful for carrying out stencil updates. Since the topology is a static description of the domain, we propose to use it to create a portable initial mapping of the location of task's data to a physical location in hardware {(i.e. the initial thread). The portable numerical identifier of the logical location of task's data is called the Software Topology Address (STA).}

Figure~\ref{fig_linear_mapping} shows an example coarsened linear mapping (e.g. Morton order) of an adaptive Cartesian mesh such as ~\ref{fig_cart_layout}, where each color represents a location (e.g. physical core). The advantage of this initial mapping is that it attempts to preserve inter and intra-task data reuse as the tasks that share a coarse key or close enough keys should be mapped to the same/nearby hardware locations.  For algebraic representations such as matrices that arise from the discretization of Partial Differential Equations (PDEs), we use the indices of the corresponding matrix blocks to create an address. The locality notion in these can be trickier as the indices of the matrix elements do not directly express topology, however, tasks that operate on the same matrix block are still guaranteed to share the same mapping. In the absence of topology, the STA is assigned to the nodes based on their relative location in the DAG (i.e. the depth of the node, and the location in the breadth). This is because it is known that nodes that are close in the DAG are much more likely to have data reuse. 
{In this case, however, the DAG should exist a-priori to auto assign the STAs based on the depth and breadth of the nodes depicted by the final DAG structure. There is no such restriction for physical domains with inherent topology structure as the STA assignment in such domains is independent of the DAG structure, so dependencies can be inserted at execution time.}
Equations~\ref{eq_sta_construction_1} to ~\ref{eq_sta_construction_4} show the formulas used to assign an initial location to a task based on its topology information, which is done in 4 stages.\\
\textbf{Stage 1: Configuring the granularity of the STA key:} since the STA is used later in this paper to index the performance model, a granularity control of how many models to create need to be available. Hence, to reduce the overhead of creating many models, $max\_bits$ is tunable parameter, which is the maximum number of bits used to express the STA.  Based on Equation~\ref{eq_sta_construction_1}, we allow $4$ times as much as there are fine-grain resource partitions (e.g. cores in the simple case). Sensitivity analysis of the granularity of STA creation is left to a future study. {Note that a ``worker'' refers to a logical thread.}
\\
\begin{equation}
\begin{split}
max\_bits &= \log_2(4 \times |workers|)\\
\end{split}
\label{eq_sta_construction_1}
\end{equation}
\textbf{Stage 2: Obtaining the key from topology:} in this stage depicted by Equation~\ref{eq_sta_construction_2}, we retrieve the space filling order ($sfo$) as an integer identifier for the model using the logical location of the task's data (e.g. Cartesian coordinates) or the location of node in the DAG if the former does not exist.\\
\begin{equation}
% \begin{split}
STA  = get\_sfo\_order(logical\_loc, max\_bits)\\
% \end{split}
\label{eq_sta_construction_2}
\end{equation}
\textbf{Stage 3: Obtaining the relative hardware location of the key:} as shown in Equation~\ref{eq_sta_construction_3}, this is obtained by dividing the STA by the maximum integer that can be represented using the defined granularity in Equation~\ref{eq_sta_construction_1}.\\
\begin{equation}
\begin{split}
relative\_loc &=  \frac{STA}{2^{max\_bits}} \\
\end{split}
\label{eq_sta_construction_3}
\end{equation}
\textbf{Stage 4: Mapping the key to initial physical location:} in the last step (Equation~\ref{eq_sta_construction_4}), we map the relative location to a worker id, which falls in a moldable resource partition as shown in Figure~\ref{fig:resource_partitions}. {The logical worker id maps to a physical location as we explain in Section~\ref{sub:mold}}
\begin{equation}
\begin{split}
worker\_id & = \lfloor relative\_loc  \times |workers|\rfloor
\end{split}
\label{eq_sta_construction_4}
\end{equation}

% Equations~\ref{eq_sta_construction_1} to ~\ref{eq_sta_construction_4} show the formula used to assign an initial location to a task based on its STA. The STA is used later in this paper to index the performance model, hence, to reduce the overhead of creating many models, $max\_bits$ is tunable parameter, which is the maximum number of bits used to express the STA. Based on Equation~\ref{eq_sta_construction}, we allow 4 times as much as there are fine-grain resource partitions (e.g. cores in the simple case) ($4 \times |workers|$). We then get the space filling order ($sfo$) to get an integer identifier for the model using the logical location (e.g. Cartesian coordinates). After that, the relative location of the partition is obtained by dividing the STA by the configured maximum integer ($2^{max\_bits}$). Finally, we obtain the partition id using the relative location and the number of workers.

% \begin{equation}
% \begin{split}
% max\_bits &= \log_2(4 \times |workers|)\\
% STA & = get\_sfo\_order(logical\_loc, max\_bits)\\
% relative\_loc &=  \frac{STA}{2^{max\_bits}} \\
% worker\_id & = \lfloor relative\_loc  \times |workers|\rfloor
% \end{split}
% \label{eq_sta_construction}
% \end{equation}
\begin{table}[t]
\caption{Example of the layout description}
\centering
% \scalebox{0.8}{
\begin{tabular}{|l|l|l|}
\hline
\textbf{Line\#} & \textbf{Content} & \textbf{Note} \\ \hline
1&0,2,4,8,1,3,5,7& Thread affinity mappings\\ \hline
2&1,2,4       &Widths for Leader thread 0\\ \hline
3&1           &Widths for Leader thread 1\\ \hline
4&1,2         &Widths for Leader thread 2\\ \hline
5&1           &Widths for Leader thread 3\\ \hline
6-9&  ...       &   ...        \\ \hline
\end{tabular}
% }
\label{table:layout_description_file}
\end{table}

\subsection{The Moldable Resource Partitioning}
\label{sub:mold}
In the case of single-threaded execution, the partition id calculated by Equation~\ref{eq_sta_construction_4} is the logical id of the smallest unit of execution exposed by the runtime environment, which is typically a hardware thread. However, to unleash the aforementioned potential of moldability, we express the system as a set of execution places.
% Elasticity refers to the ability of the system to match the assigned resources to the workload requirements. 
The system in Figure~\ref{fig:resource_partitions} shows an example hardware configured with 2 places with $W=4$, 4 places with $W=2$ and 8 places with $W=1$. A place encompasses the cores that share a resource (e.g., cache level, memory subsystem, network element). Thus, the relative location resembled by the line segments below Figure~\ref{fig:resource_partitions} and indicated by Equation~\ref{eq_sta_construction_3} points to a moldable resource partition (initially, $W=1$ is selected). To allow the flexible allocation of resources, the scheduler optionally accepts a layout description file at initialization phase. For example, the dual-socket 4-core system shown in Figure~\ref{fig:resource_partitions} can be described using the file whose content is highlighted in Table~\ref{table:layout_description_file}. {Line\#1 of Table~\ref{table:layout_description_file} represents the thread affinities (hardware thread id) for each of the logical threads (workers). So in this case, the runtime is configured with 8 workers, which are mapped to the listed hardware thread ids. For example worker 0 is mapped to thread id 0, worker 1 is mapped to 2 and so on. In the following 8 lines, the supported resource widths for each of the workers is specified. Hence, for leader worker 0, the supported widths are 1, 2 (spanning workers 0 and 1), and 4 (spanning workers 0 - 3).} This flexible specification can easily support heterogeneous architectures, however, the evaluation of \sscheduler\/ on such platforms is currently work in progress. Note that there exist tools  such as  \texttt{hwloc}{~\cite{hwloc}} that can realize the system layout, and thus can be used here to generate the layout description.  
\subsubsection{Moldable work-stealing}

\label{sub:moldable_ws}
\begin{figure}
    \centering
    \efbox{
    \includegraphics[width=.7\columnwidth]{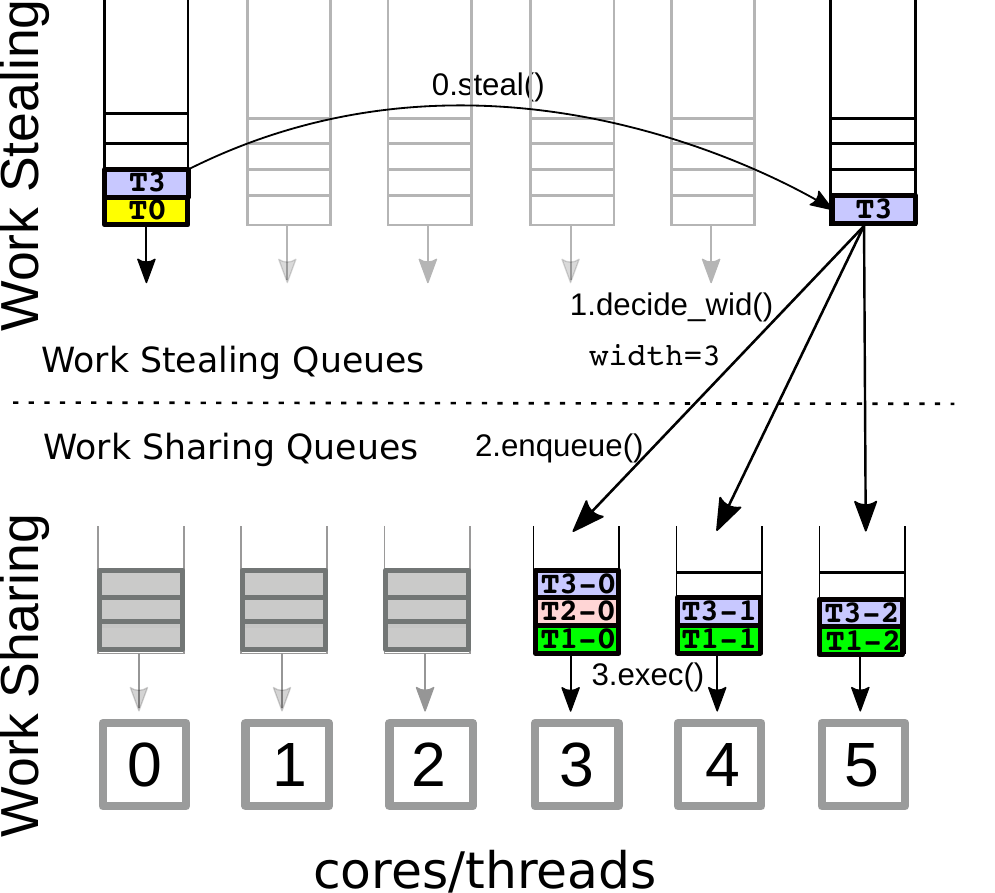}}
    \captionof{figure}{Moldable work-stealing scheduler}
    \label{fig:moldable_ws}
\end{figure}

{Since \sscheduler\/ is based on work-stealing, we present Figure~\ref{fig:moldable_ws} to show the interaction between work-stealing and work-sharing queues for moldable task execution. Each worker thread has a work-stealing and a work-sharing queue. The work-stealing queue includes the tasks that either are initially assigned the thread (i.e. the STA mapped thread), or the tasks that are stolen to achieve a load-balanced execution (work-balancing modes are covered in Sections~\ref{sec:locality_scheme} and~\ref{sec:workbalancing_scheme}). In Figure~\ref{fig:moldable_ws}, \texttt{T3} reaches the front of the queue. At this stage, the width is decided using the \sscheduler\/ performance model, then the task is inserted to the work-sharing queues of the execution place. The queues are based on a lock-free implementation adopted from~\cite{lock-free-Sutter}. So \texttt{T3} is assigned to the partition $R=[LR=3,W=3]$ which represents the partitions (\texttt{T3-0, T3-1, T3-2}). These are executed asynchronously on workers (3,4,5). The task work function is entitled to implementing the internal scheduling scheme (similar to the OpenMP parallel regions). Hence, each task receives a partition id in the range [0,width-1], and has access to $R$. In this paper, we adopt a static scheduling scheme to preserve thread-level data reuse across dependent tasks. This is semantically similar to OpenMP's static schedule for loop-parallel codes. } 

\begin{algorithm}
 \SetAlgoLined
 \SetKw{KwGoTo}{go to}
 \While{active}{
  \uIf{!local\_queue.empty()}{
  task = pop\_local\_queue()\; \label{alg:line_start_loc}
  partitions = inclusive\_partitions[local\_th\_id]\;\label{mold}
  res\_part = min\_cost\_part(task.type, task.sta, partitions)\;
  time = exec(task, res\_part)\; \label{exec}
  cost = time.leader x res\_part.width\;
  update\_cost\_part(task.type, task.sta, res\_part);\label{alg:line_end_loc}
%   \If{cost < min\_cost(task.type, task.sta, partitions)}{
%     update\_min\_cost\_part(task.type, task.sta)\;\label{alg:update_min_cost}
%   }\label{alg:line_end_loc}
  }\uElseIf{!inc\_part\_queues.empty()}{
    task = pop\_one\_inc\_queue()\; \label{alg:line_start_inc}
     \KwGoTo \ref{mold}\;\label{alg:line_end_inc}
  }
  \uElseIf{!non\_inc\_part\_queues.empty()}{
 \label{alg:line_start_global}
    \uIf{stealing\_attempts > THRESHOLD}{\label{alg:very_idle}
     task = pop\_non\_inc\_rand\_queue()\;
     \KwGoTo \ref{mold}\;
    }\Else{
    task = peak\_non\_inc\_rand\_queue()\;
    partitions = get\_all\_partitions()\;\label{alg:line_get_global_min_1}
    res\_part = min\_part\_cost(task.type, task.sta, partitions)\;\label{alg:line_get_global_min_2}
    \If{local\_th\_id in res\_part} {\label{alg:has_better_partition}
      task = pop\_non\_inc\_rand\_queue()\; 
      \KwGoTo \ref{exec}\;
    }  \label{alg:line_end_global}
    }
   }
 }
 \caption{\sscheduler\/ work and resource selection}
 \label{alg:lagres}
\end{algorithm}
\subsection{The Online Performance Model}
\label{sub:online_perf}
\textbf{Identifying the model:}
A key contribution of this work is defining the parameters that constitute a locality-adaptive performance model for a task. Without STA as a key (Figures~\ref{fig_cart_layout} and ~\ref{fig_linear_mapping}), the interpretation of a model would neglect the effect of data locations.  
Since the STA implies the mapping of task's data to a hardware location, it facilitates creating a performance model per locality for a specific task. This is to study the effects of preserving data locality for that task. Once it is initialized (e.g. using NUMA first touch policy or \texttt{hwloc-lib}'s explicit memory pinning) in a resource partition $R_i=[LR=j,W=1]$, the model studies the effect of molding the task and its dependencies that have the same work function. In our implementation, we express tasks as object-oriented classes for easier \texttt{C++} type resolution using \lstinline{std::type_id} at runtime. Alternatively, a distinct integer for each task's work function can be used for non Object-Oriented specifications. 
Adding the STA enables to model the performance per locality on the available system partitions. This manifests as a 2D array structure (\lstinline{model[type_index][sta]}). A reference to the performance table is assigned to each task at initialization phase. For each model (Figure~\ref{fig_perf_model}), the following schemes are studied:
\begin{itemize}[leftmargin=*]
    \item The locality scheme: for the thread's local tasks, schedule within the set of inclusive partitions of the initial thread, i.e., all the partitions that share a resource with the initial task's thread as depicted by the layout description. For locality sensitive tasks, a cost-efficient place can belong to this set. 
    \item The work-balancing scheme: use if thread's local queue is empty.
\end{itemize}
\textbf{The used performance modeling scheme:}
We adopt an online history-based scheme 
% {, 
% which is an extension of Estimated Execution Time~\cite{son_efficient_2013} and is 
% } 
similar to that used by \texttt{StarPU} \cite{starpu-model-augunnet-europar-2009} task programming library. We extend it to model the performance on execution places. This renders it  powerful in accurately predicting the performance of fine-grain tasks, which are typical since one of the important purposes of fine-grain task-parallelism is maximizing concurrency by creating many tasks (Task Count $\gg$ Execution Places). Therefore, the history-based scheme proves to be effective as we show in Section~\ref{sec:eval}. However, the implementation of the performance model is decoupled from the scheduler. Thus, models such as regression based or analytical models can be seamlessly used in \sscheduler. 
We start by training the model on the available places as the number of tasks that share a model is much larger than the available places (a simple recursive task parallel code like \texttt{SparseLU} on 64x64 blocks results in 12k tasks $\gg$ partitions).~{The online history-based scheme assumes that performance is insensitive to the DAG iteration number, so a previous iteration's value can be used to predict the current iteration. However, the dynamic changes in performance can be detected as the timing values are continuously updated for the selected execution places as we show in Section~\ref{sec:locality_scheme}.} 
% However, linear and non-linear regression-based approaches can be used to reduce training time if it is observed to pose an overhead. 
Eventually, for the model to result in an efficient mapping (Figure~\ref{fig_task_mapping}) the locality and the work-balance schemes are used as shown in Sections~\ref{sec:locality_scheme} and ~\ref{sec:workbalancing_scheme}. \\
{
\textbf{On the runtime overhead of the training scheme:}
The initial STA mapping determines the distribution of the tasks over the worker threads. This offers a way to balance the work since the STA identifiers are spread over the input domain. For example, the space-filling order mapping is traditionally used to evenly partition domains with known input coordinates. Also, the even distribution is also achieved when STAs are automatically assigned by the runtime. The STAs are then normalized to the available threads as we show in Section~\ref{sub:sta}. This reduces the penalty of under-utilization due to imbalanced partitioning. \newline
In the case of the history-based model adopted here, the timetable is greedily filled in increasing order of the resource width (i.e. starting the execution from $W=1$). 
The locality scheme highlighted in Section~\ref{sec:locality_scheme} guarantees that the initial thread is always included in the subsequent schedules of the task ($W\geq1$), which ensures producer-consumer data reuse between dependent tasks.
We note that, since we use an online approach, we do not separate the training phase from the actual execution of the application; all work contributes to training, which does not have the overhead imposed by offline schemes. Based on our profiles, the impact of timetable filling and the sub-optimal resource width choices is negligible as we start from the assumption that Task Count $\gg$ Execution Places, which is valid across all experiments in Section~\ref{sec:eval}.
Nonetheless, if this assumption does not hold (e.g. in the case of relatively many execution places), the layout description (e.g. Table~\ref{table:layout_description_file}) can be adjusted such that the requested partitions do not span beyond the resources needed by the task with the max working set.  
% Section~\ref{sec:locality_scheme} shows how we overcome the known issues of over-utilization or under-utilization of resources due to exhausting scheduling options at training phase. 
}
\subsubsection{The locality scheme}
\label{sec:locality_scheme}
% For latency-bound workloads similar to Figures~\ref{fig:copy_small_abu_dhabi}~\ref{fig:copy_large_abu_dhabi}, mold
The purpose of this scheme is to adjust the resource width within the inclusive sharing partitions. 
As mentioned before, a moldable task involves a work-sharing region that execute on a resource partition constituting one or multiple workers executing asynchronously. 
Table~\ref{tbl:local_model} shows the local partitions spanning threads 0 - 3 of the system shown in Figure~\ref{fig:resource_partitions}. {This means that logical thread 0 is included in the local resource partitions ($LR=0, W=1$), ($LR=0, W=2$), and ($LR=0, W=4$)}. When a task is initialized in a given thread (as per the STA), the modeled performance (i.e. execution time herein) of the inclusive partitions (e.g. Table~\ref{tbl:local_model}) is fetched. {This serves as a lookup table for the indices of the local partitions of each thread. It is initialized at startup given a layout description file such as Table~\ref{table:layout_description_file}}. Hence, to predict the effect of moldability (i.e. changing the resource width), the partition that minimizes the cost function $f(LR=j, W=w) = T(j) \times w$ is selected for scheduling. 
This cost is depicted by the cpu time perceived by the leader thread ($T(LR)$) multiplied by the resource width ($W$). {Therefore, the higher the parallel cost, the lower the selected width and vice versa. The new cost is updated as per the new cost (Line~\ref{alg:line_end_loc}).}
Eventually, for a latency-bound workloads, the model picks the width that matches the task's working set size and minimizes the resource over-subscription.
Also, in the events of lower DAG parallelism (at coarsening phases), the parallel cost will be lower, since more workers are available to execute the task. This results in the task being dynamically mapped to {a local partition of width $w$ that increases utilization}.
% Here, the selected width tends to be smaller than the larger sized input, where larger widths are deemed more efficient.
Algorithm~\ref{alg:lagres} (Lines~\ref{alg:line_start_loc} -~\ref{alg:line_end_loc}) highlights the locality scheme of \sscheduler.
% % \hspace*{-0.3cm}
% \begin{minipage}[t]{0.5\columnwidth}
% % \begin{figure}
%     \centering
%     \includegraphics[width=1.0\columnwidth]{figs/local_model.pdf}
%     \captionof{figure}{Parallel cost analysis}
%     \label{fig:local_model}
% % \end{figure}
% \end{minipage}
% % \rulesep 
% \hspace{0.2cm}
% \begin{minipage}[t]{0.5\columnwidth}
% \vspace{-1.35cm}
% \scalebox{0.7}{
% \begin{tabular}{|c|c|}
% \hline
%  Init thread & Local inclusive partitions  \\
% \hline
%  0 & $\{[0,1],[0,2],[0,4]\}$  \\
% \hline 
%  1 & $\{[1,1],[0,2],[0,4]\}$  \\
% \hline
%  2 & $\{[2,1],[2,2],[0,4]\}$  \\
% \hline
%  3 & $\{[3,1],[2,2],[0,4]\}$  \\
% \hline
% \end{tabular}
% }
%     \captionof{table}{\\Modeled inclusive partitions}
%     \label{tbl:local_model}
% \end{minipage}
% \begin{figure}
%     \centering
%     \includegraphics[width=0.6\columnwidth]{figs/local_model.pdf}
%     \captionof{figure}{Parallel cost analysis}
%     \label{fig:local_model}
% \end{figure}
\begin{table}
    \caption{Modeled inclusive partitions}

\begin{tabular}{|c|c|}
\hline
 Init thread & Local inclusive partitions  \\
\hline
 0 & $\{[0,1],[0,2],[0,4]\}$  \\
\hline 
 1 & $\{[1,1],[0,2],[0,4]\}$  \\
\hline
 2 & $\{[2,1],[2,2],[0,4]\}$  \\
\hline
 3 & $\{[3,1],[2,2],[0,4]\}$  \\
\hline
\end{tabular}

    \label{tbl:local_model}
    \end{table}
% \hspace*{-0.3cm}
% \begin{minipage}[t]{0.5\columnwidth}
% % \begin{figure}
%     \centering
%     \includegraphics[width=1.0\columnwidth]{figs/local_model.pdf}
%     \captionof{figure}{Parallel cost analysis}
%     \label{fig:local_model}
% % \end{figure}
% \end{minipage}
% % \rulesep 
% \hspace{0.2cm}
% \begin{minipage}[t]{0.5\columnwidth}
% \vspace{-1.35cm}
% \scalebox{0.7}{
% \begin{tabular}{|c|c|}
% \hline
%  Init thread & Local inclusive partitions  \\
% \hline
%  0 & $\{[0,1],[0,2],[0,4]\}$  \\
% \hline 
%  1 & $\{[1,1],[0,2],[0,4]\}$  \\
% \hline
%  2 & $\{[2,1],[2,2],[0,4]\}$  \\
% \hline
%  3 & $\{[3,1],[2,2],[0,4]\}$  \\
% \hline
% \end{tabular}
% }
%     \captionof{table}{\\Modeled inclusive partitions}
%     \label{tbl:local_model}
% \end{minipage}

\subsubsection{The work-balancing scheme}
\label{sec:workbalancing_scheme}
Work-imbalance in dynamically scheduled DAGs is a well-studied problem, { and it is especially considered in task-based runtime systems}. One of the known solutions is treating the side effect of imbalance (i.e. idleness) using distributed work-stealing. In this approach, each worker thread has its own local queue that can be stolen from in events of idleness based on non-deterministic/random decisions ~\cite{opt-dist-work-stealing-kumar-iaaa16,dist-work-stealing-paudel-icpp13} or 
deterministic decisions~\cite{shumpei-sc19}. \sscheduler\/ uses distributed work-stealing queues. However, the stealing decisions are not entirely random. \sscheduler\/ follows a heuristic method to try to maximize locality, based on the steps below:
\begin{itemize} [leftmargin=*]
\item Local work-stealing (Algorithm~\ref{alg:lagres} (Lines~\ref{alg:line_start_inc} - ~\ref{alg:line_end_inc})): this is done by checking the queues of the threads of inclusive partitions. For example, if thread id $3$ is idle, then based on Table~\ref{tbl:local_model}, the inclusive threads are $0,1,2$. The queues of these candidates are checked in round-robin fashion starting from the $(thread\_id+1)\%{inc\_set\_size}$ \footnote{This step is dropped from Algorithm~\ref{alg:lagres} for brevity}. When a task is found, similar cost analysis to the locality scheme is applied. 
\item Non-local work-stealing (Algorithm~\ref{alg:lagres} (Lines~\ref{alg:line_start_global} - ~\ref{alg:line_end_global})): if no work is found from previous step, a random non-local task is fetched. The resource partition that globally minimizes the cost is fetched (Lines~\ref{alg:line_get_global_min_1} - ~\ref{alg:line_get_global_min_2}). The stealing thread checks if it falls in the partition of the global minimum, otherwise, the stealing attempt is rejected (i.e. when condition in Line~\ref{alg:has_better_partition} evaluates to false). When the attempts reach certain threshold, the request will be fulfilled anyway (Line~\ref{alg:very_idle}).
     
\end{itemize}

% \subsection{The Scheduling Algorithm}
% \label{sec:alg}

% The \scheduler tries to achieve 

% \input{implementation}

% \input{methodology}
\begin{table*}
\caption{Hardware structure of the used machines. Surrounded by parenthesis is the number of hardware threads that share memory/cache.}
\centering
\begin{tabular}{c|c|c|c|c|c|c|c}
\hline
Architecture& Sockets  & Cores/socket & Threads/core &Memory(GB) & L3(MB) & L2(KB) & L1d(KB) \\
\hline
\intelskl\/~\cite{intel-skylake-specs} & 2 & 16 & 1 & 48(16) & 22(16) & 1024(1) & 32(1) \\
\hline
\end{tabular}
\label{tab:hw}
\end{table*}

\section{Methodology}
\label{sec:methodology}
This section describes the experimental methodology used to evaluate the contributions of this works. \sscheduler\/ is integrated into \xitao\/ ~\cite{xitao},
a DAG runtime system implemented on top of modern C++ threading extensions ($\geq$ \texttt{c++11}). \xitao\/ is designed to flexibly evaluate scheduling policies and already features moldable tasks. 
% This facilities the mapping of $N$ tasks to $M$  resources~\cite{blind-reference}. 
However, \sscheduler\/ is decoupled from the runtime internals. 

\subsection{Platforms}
Experiments are performed on an \intelskl\/ (Intel Xeon Gold 6130), which is a modern multicore architecture with the memory hierarchy described in Table~\ref{tab:hw}. The nodes are hosted at Tetralith (NSC's largest HPC cluster), which consists of 1908 compute nodes each with a dual socket 16-core \intelskl CPU, giving a total of 61056 CPU cores. \xitao\/ is configured to run with 32 worker threads\\
% One node on each machine is used exclusively (i.e. with no other jobs co-scheduled on the node during the experiments).\\
\textbf{The layout description file}: worker threads are mapped to the physical cores of the platform totaling 32 cores. The configured resource widths are 1, 2, 4 and 16. This means that we do not map a  task across the 2 sockets. {Note that there is no restriction on mapping a task to two sockets. Based on our traces, the scheduler will directly identify this as a suboptimal choice due to the cost of NUMA misses resulting from accesses from remote workers. This reflects as a high modeled parallel cost.} From a memory perspective, a task can leverage (1, 2, 4, 16) x L1/L2 caches and the L3 cache, with width 16.

% \intelskl\/ is hosted at the Tetralith 
% The \amdad\/ machine is a 4 AMD Opteron processors (6348), each has 2 chips, with 6 cores for each chip, for a total of 48 cores.
% The other two machines belong to academic research computing centers:[omitted for blind review]. One node on each machine is used exclusively (i.e. with no other jobs co-scheduled on the node during the experiments).
% The nodes of the \intelhw\/ cluster have 2 Xeon E5-2650v3 processors, with 10 cores each, for a total of 20 cores.
% Also, we use a ``large memory'' \intelbw\/ cluster, where each node contains 4x18 Xeon E7-8860v4 cores, for a total of 72 cores.

\subsection{Baseline Schedulers}
Below we describe and justify the baseline schedulers used:\\
\textbf{Random Work-Stealing Scheduler (\rwss\/)}: it is based on distributed work-stealing, where each worker greedily tries to reduce idleness by fetching work from victim queues. It has been formally introduced in~\cite{blumofe-jacm99} and has appeared in several parallel task-based libraries such as Cilk~\cite{blumofe-cilk}, %MassiveThreads~\cite{nakashima-lncs14} %
and Intel TBB~\cite{kukanov-itj11}.\\
\textbf{Almost Deterministic Work Stealing (\adws\/)}: this scheduler is introduced in~\cite{ShiinaSC19_ADWS}. It is currently the state-of-the-art in work-balanced locality-aware scheduling developed as an extention to the scheduler of the MassiveThreads library~\cite{Nakashima2014_Massive_threads}. It follows a deterministic task allocation based on programmer workload hints, where the work is recursively split across the DAG nodes (spawned and continuation tasks). Each worker maintains a local and a migration queue, and work stealing is only allowed inside the ``work groups''. A port of \adws\/ has been implemented in \xitao\/.\\
\textbf{\lags\/}: We also evaluate the proposed scheduler using 1:1 task mapping. %without the effect of moldability. 
Partition widths are persistently set to 1. Tasks are initialized and executed in the NUMA node mapped by their STA. Local stealing is preferred, then global stealing requests are fulfilled when the stealing thread reaches idleness threshold and the steal that reduces the cost function (as per the model) is chosen. \\
\textbf{\armsn\/}: The proposed scheduler with all components from Section~\ref{sec:lagres}, using 1:M task mapping. The scheduler is configured with the  parameters shown in Table~\ref{table:scheduler_params}:

\begin{table}
\caption{The configuration parameters to enable \sscheduler\/ in \xitao\/}
% \scalebox{0.9}{
\begin{tabular}{@{} *5l @{}}    \toprule
\emph{Option} & \emph{Value} & \emph{Usage}  \\\midrule
\texttt{sta}    & enabled  & capture/assign sta values        \\ 
\texttt{perf-model}    & enabled  & build online perf model        \\ 
\texttt{idle-tries}    & 10  & idle iterations before stealing        \\ 
 \texttt{local-steal} & enabled & use local balancing scheme \\
\texttt{global-steal} & enabled & use non-local balancing scheme  \\
\texttt{moldablity} & enabled & map a task to M workers  
 \\\bottomrule
 \hline
\end{tabular}
% }
\label{table:scheduler_params}
\end{table}
\subsection{Synthetic Benchmark}
\begin{figure}[t]
    \centering
    \includegraphics[width=0.5\columnwidth]{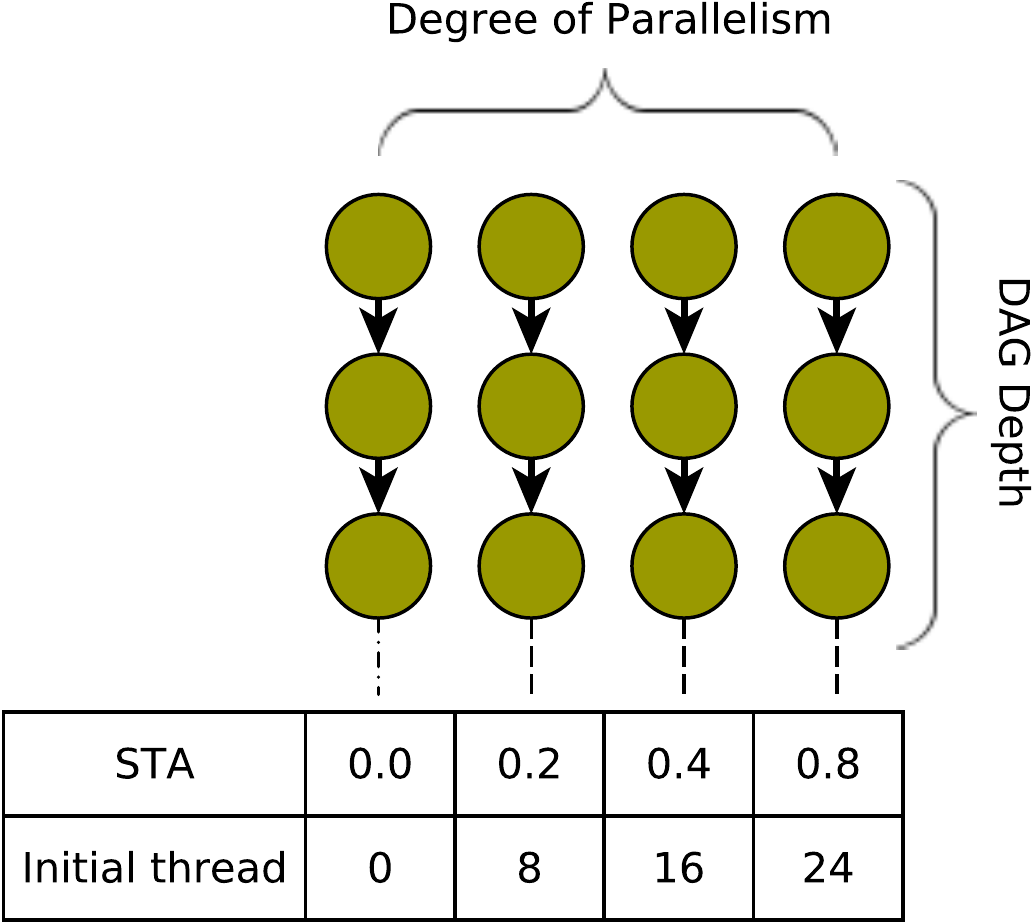}
    \caption{The synthetic benchmark used to evaluate \sscheduler\/. This example assumes 32 threads. Number of Tasks = Parallelism $\times$ Depth}
    \label{fig:synthbench}
\end{figure}
To validate the adaptive online performance model adopted by \sscheduler\/, we use a synthetic DAG benchmark as shown by Figure~\ref{fig:synthbench}. The degree of DAG parallelism as well as the depth of the DAG are control parameters. This is useful to fix the number of tasks and understand the trade-off between parallelism and locality. Also, the tasks can be configured to be either matrix multiplication (MatMul) or Stream Triad tasks. {For each chain, the table underneath Figure~\ref{fig:synthbench} shows the relative STA location of the thread obtained from Eq~\ref{eq_sta_construction_3}. The initial thread is acquired using  Eq~\ref{eq_sta_construction_4} with the 32 worker threads available in the experimental platform. }
\begin{figure}[]
\centering
\subfigure[Stencil 2D DAG with timesteps = 2]{ \includegraphics[width=0.42\columnwidth]{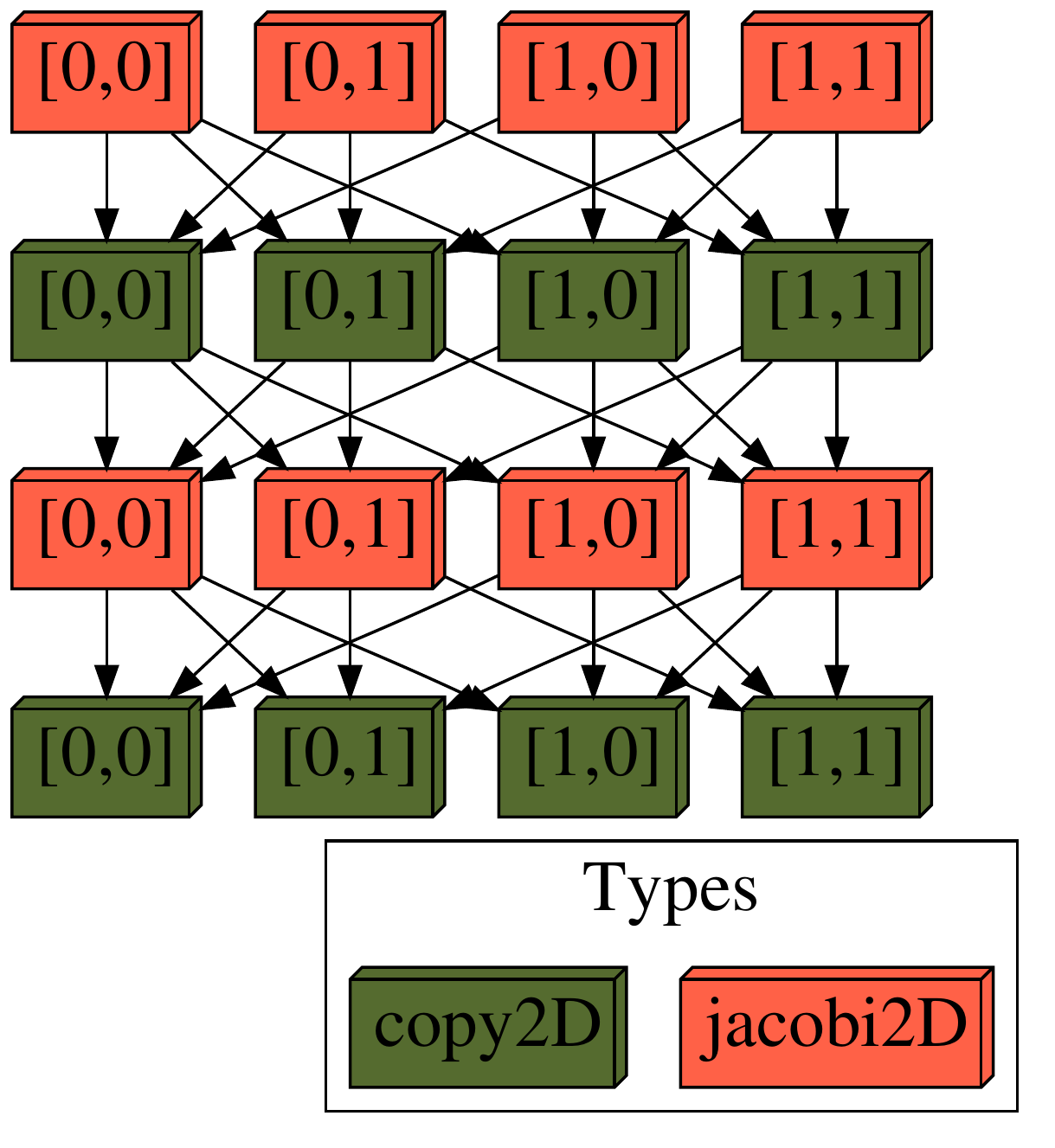}
\label{fig_dag_heat}}\hspace{2pt}
% \rulesep
\subfigure[SparseLU DAG for 4x4 blocks]{ \includegraphics[width=0.42\columnwidth]{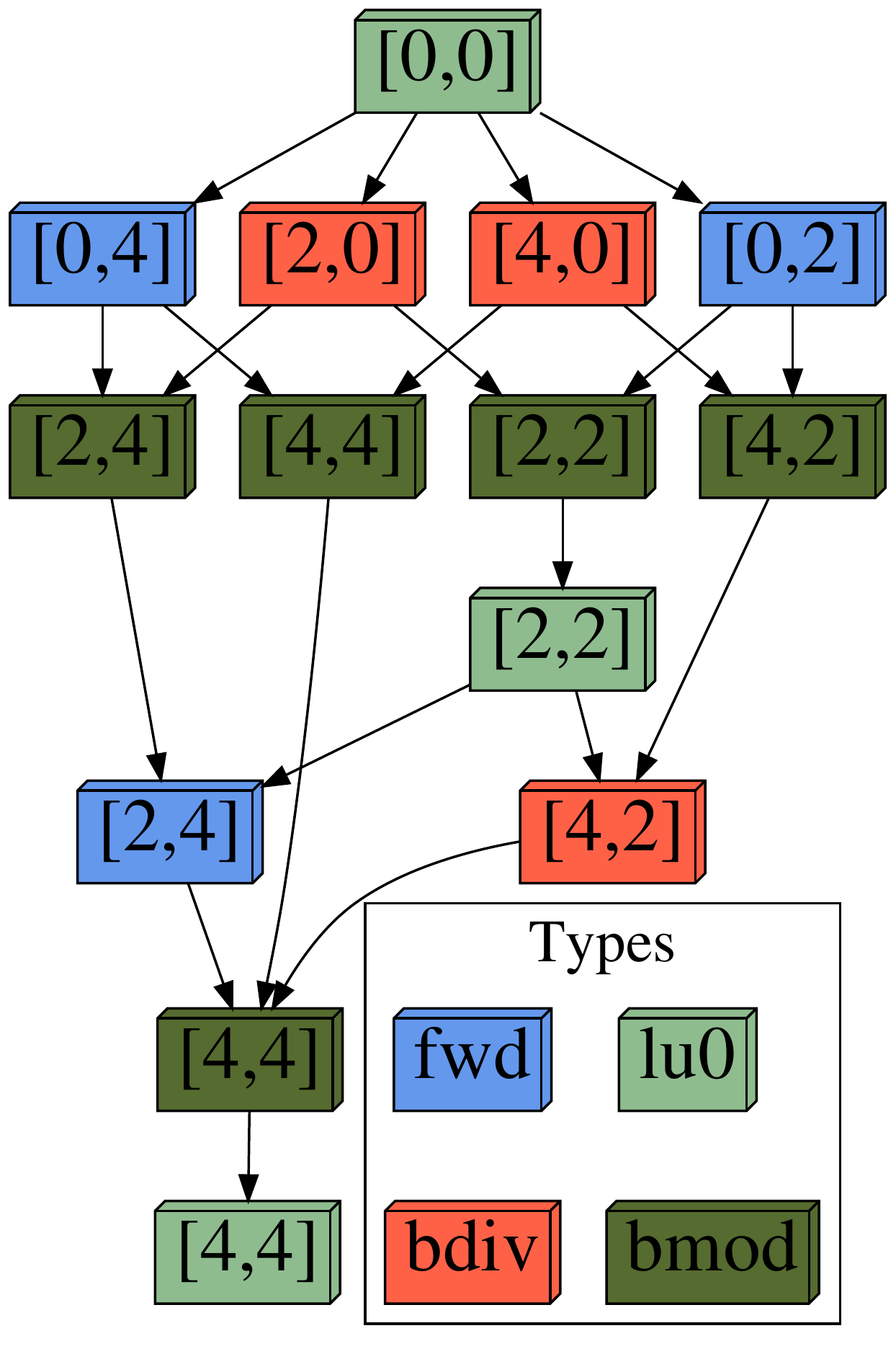}
\label{fig_dag_sparselu}} \\
\subfigure[FMM DAG Overview]{ \includegraphics[width=0.42\columnwidth]{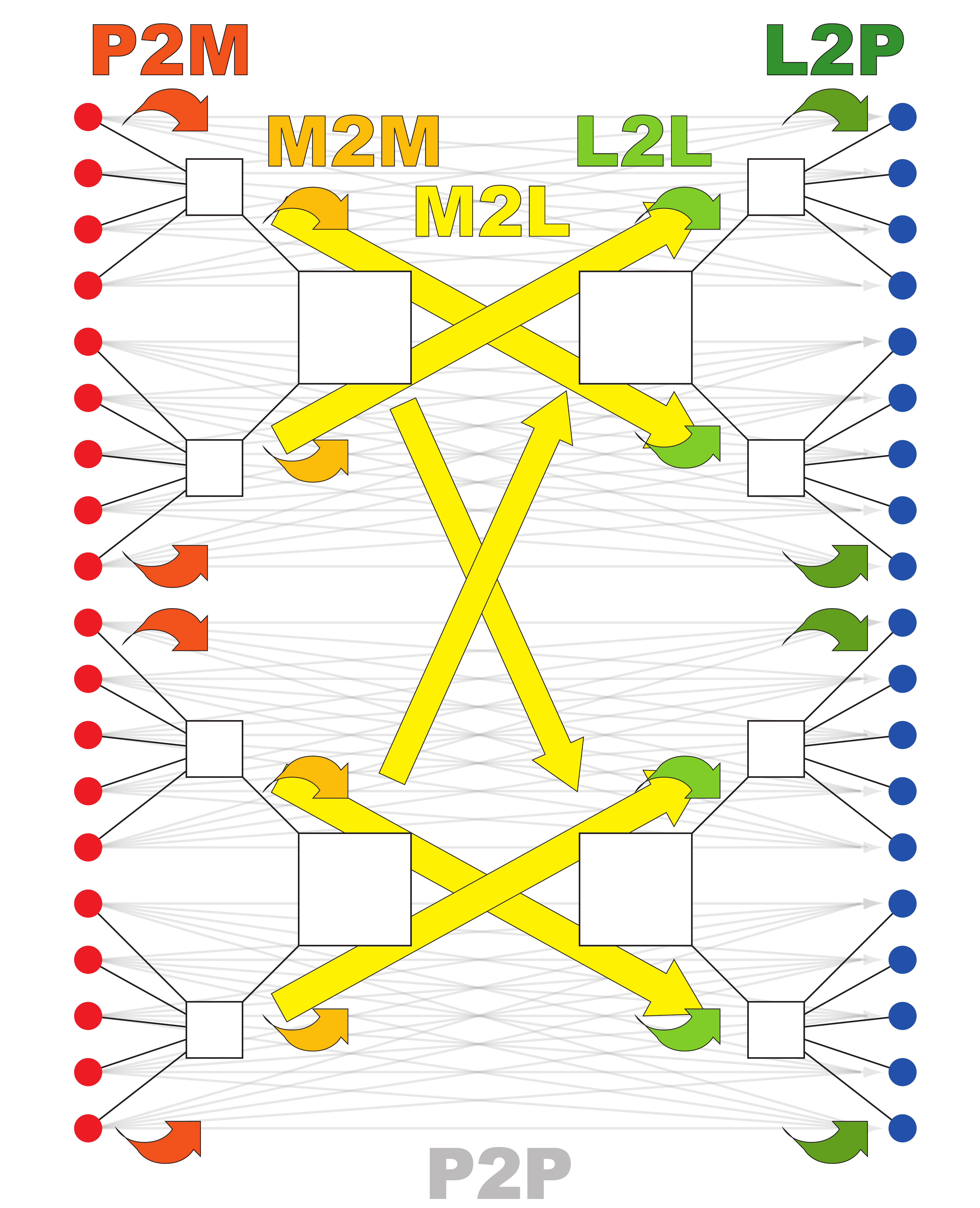}
\label{fig_fmm}} \hspace{2pt}
\subfigure[MatMul DAG with 2 levels of Recursion]{ \includegraphics[width=0.42\columnwidth]{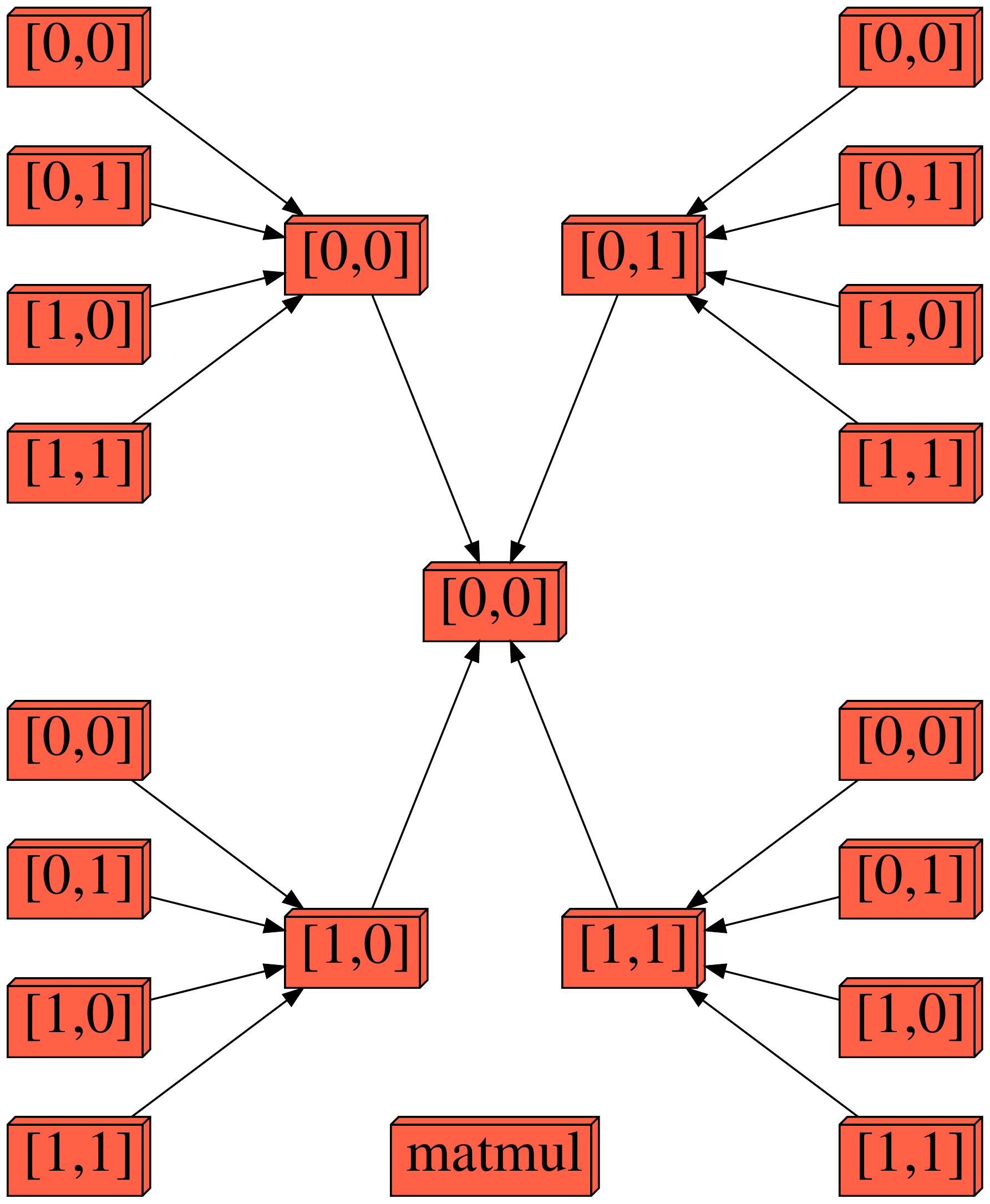}
\label{fig_dag_matmul}} 
\caption{Sample dependency graphs for the analyzed applications.}
\label{fig_app_dags}
\end{figure}
\subsection{Applications}
\textbf{Iterative DAG - HEAT:} we leverage a DAG implementation to compute heat diffusion on a 2D grid. One of the iterative numerical methods to achieve this is to use 2D Jacobi stencil. We use a 5-point stencil and create dependencies between the neighbor nodes (see Figure~\ref{fig_dag_heat}). The approach involves computing the stencil in a \texttt{compute} task, and copying out the update in a \texttt{copy} task. The DAG is iteratively executed  for a fixed number of 2k iterations. \textit{For STA specification, we use the coordinates of block of mesh points involved in a task.}\\
\textbf{Recursive DAG - SparseLU:} we port a SparseLU benchmark from the Barcelona OpenMP Tasks Suite~\cite{duran-icpp09}. This benchmark computes an LU matrix factorization over sparse matrices. The matrix is composed of NxN blocks, each of which has a pointer to a sub-matrix of size MxM. Load-imbalance is evident due to the sparsity of the matrix. For each phase of the LU algorithm, a task is spawned for non-empty blocks (see Figure~\ref{fig_dag_sparselu}). \textit{For STA specification, the matrix block indices are used.} \\
\textbf{Recursive DAG - The Fast Multiple Method (FMM):} we port the \texttt{fmm-minimal} task-based implementation from the exafmm library~\cite{exafmm-minimal} to \xitao\/. FMM is a popular $O(N)$ solver for the matrix vector products arising from the solution of certain boundary integral equations \cite{abduljabbar_sisc19_bemfmm}.
It is originally used to efficiently solve the quadratic $N$-Body problem that appears from particle/gravitational simulations (e.g. from molecular dynamics and astrophysics). 
This implementation is tree-based (see Figure~\ref{fig_fmm}). We set the leaf cell size to 64 particles.\textit{ The STA in this case leverages the Cartesian coordinates of the underlying tree cell.}\\
\textbf{Recursive DAG - MatMul:} this is a cache-oblivious divide-and-conquer \cite{cache-oblivious-frigo} implementation of the dense matrix multiplication. It is favorable to task-parallelism since the recursive subdivision to smaller blocks can be assigned to tasks with respective dependencies (see Figure~\ref{fig_dag_matmul}). The block size for the smallest block has been set to 128-256 across all experiments. \textit{The STA is the block indices per each level of the recursion tree. }

% \input{evaluation_new}
% \begin{table*}
% \caption{Hardware structure of the used machines. Surrounded by parenthesis is the number of hardware threads that share memory/cache.}
% \centering
% \begin{tabular}{c|c|c|c|c|c|c|c}
% \hline
% Architecture& Sockets  & Cores/socket & Threads/core &Memory(GB) & L3(MB) & L2(KB) & L1d(KB) \\
% \hline
% \amdad\/~\cite{amd-abu-specs} & 4 - 2 nodes each & 6 & 2 & 8(6) &6(6) & 2048(2) & 16(1) \\
% \intelbw\/~\cite{intel-broadwell-specs}  & 4 & 18 & 1 & 768(18) &45(18) & 256(1) & 32(1) \\
% \intelhw\/~\cite{intel-haswell-specs} & 2 & 10 & 1 & 256(10) &24(10) & 256(1) & 32(1) \\
% \hline
% \end{tabular}
% \label{tab:hw}
% \end{table*}

\begin{figure*}[t]
\subfigure[Parallel chains of Compute Intensive Matmul Tasks - Number of Tasks = 50k. N = 128 per task]{
\includegraphics[width=0.3\textwidth]{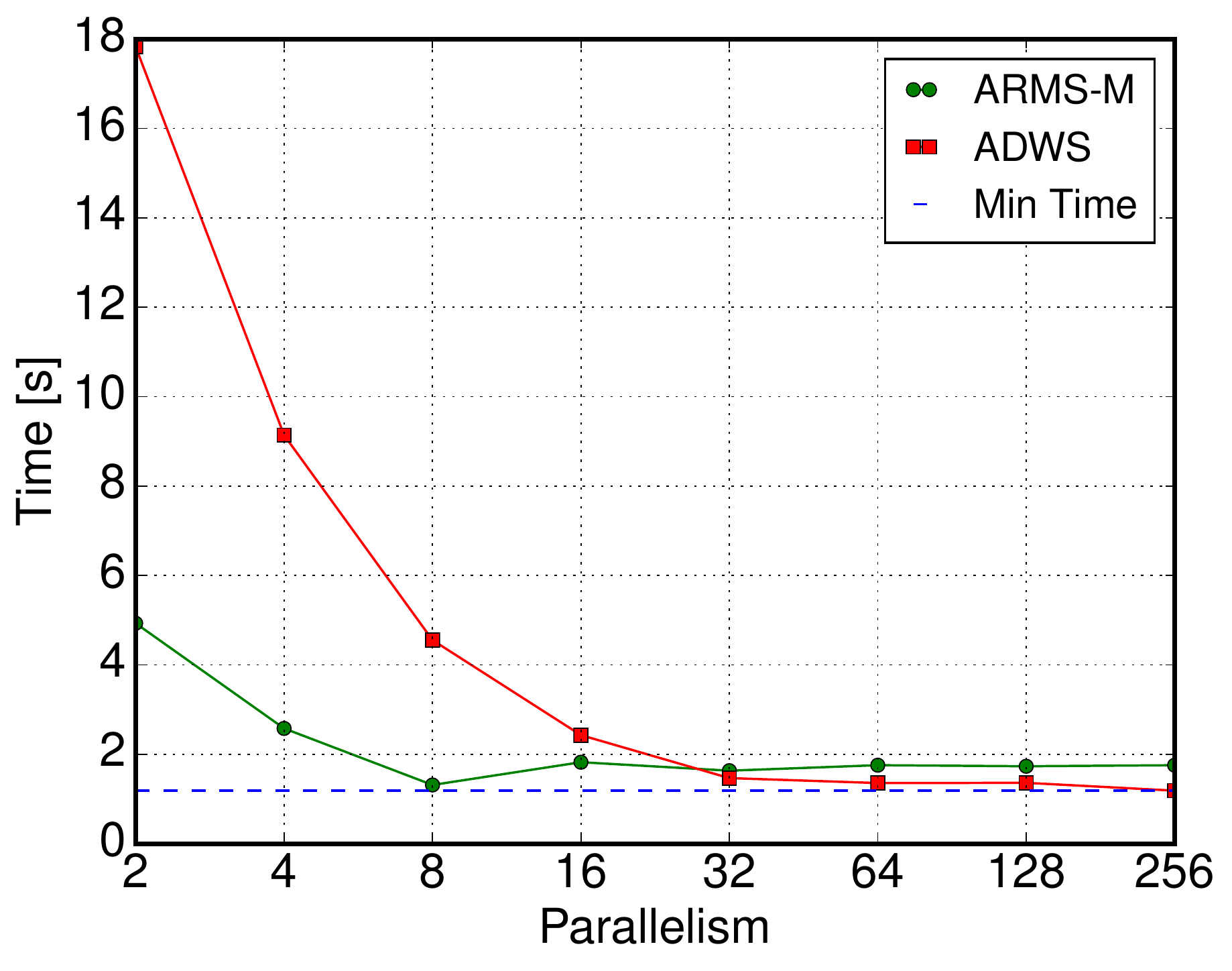}
\label{fig:chain_compute_locality_parallelism}
}
\subfigure[Parallel chains of Memory Intensive Triad Tasks - Number of Tasks = 50k. N = 512 per task]{
\includegraphics[width=0.3\textwidth]{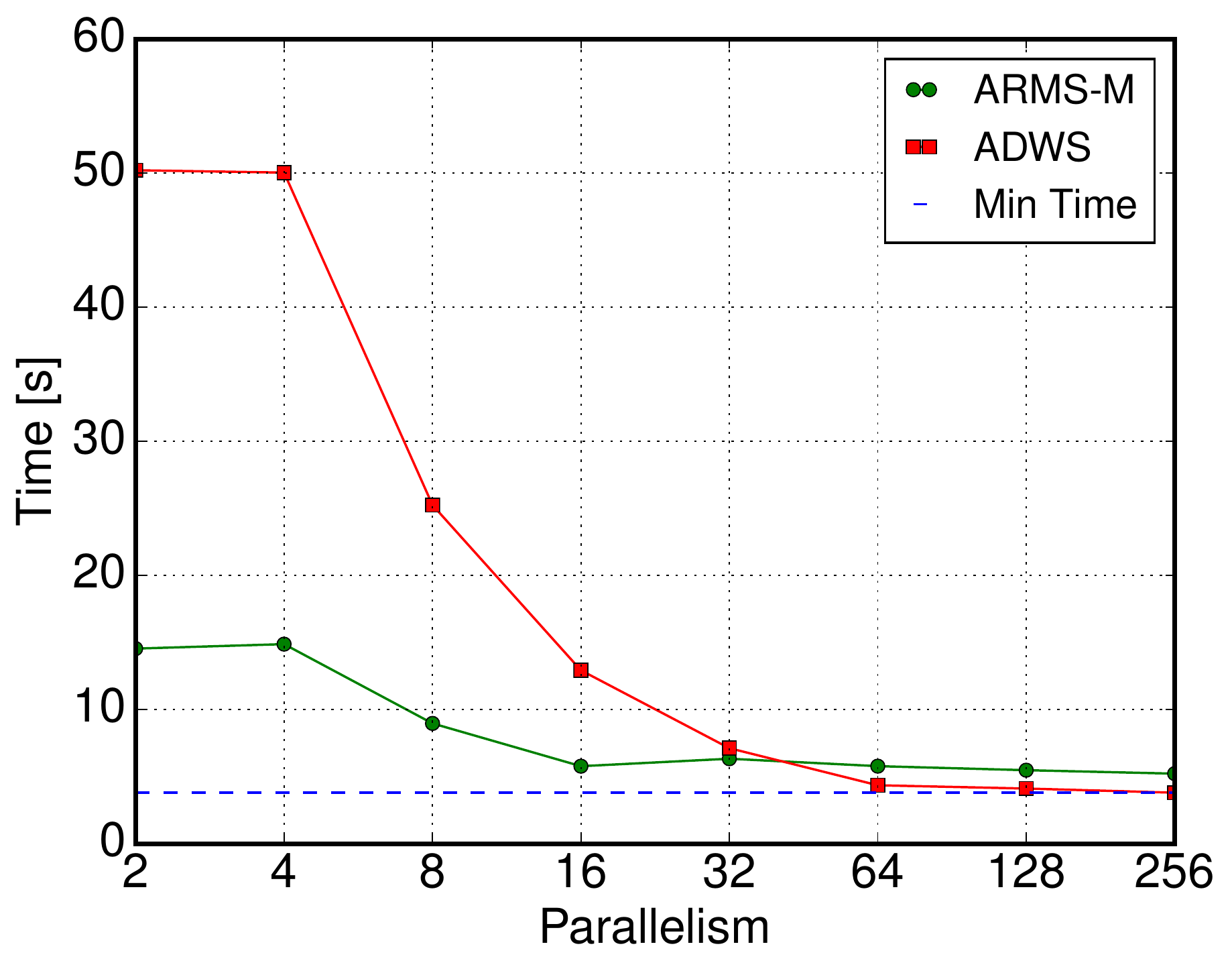}
\label{fig:chain_memory_locality_parallelism}
}
\subfigure[Parallel chains of mix of Triad and Compute Tasks - Number of Tasks = 25k Triad and 25k Matmul tasks. N = 512 per task]{
\includegraphics[width=0.3\textwidth]{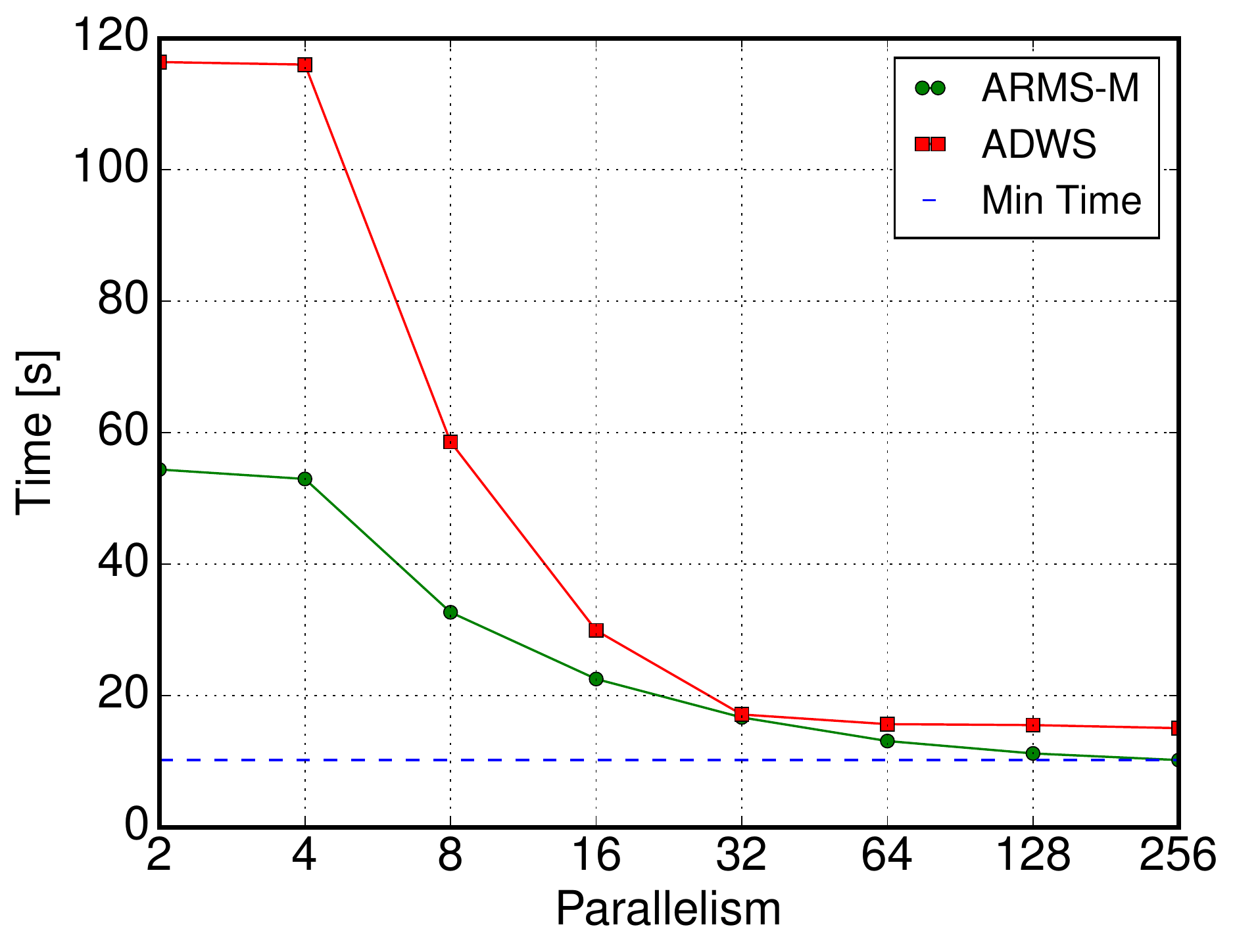}
\label{fig:chain_mixed_locality_parallelism}
}

\caption{\sscheduler\/ performance at different DAG parallelisms. }
\label{fig:chain_locality_parallelism}
\end{figure*}

\section{Evaluation}
\label{sec:eval}
In this section, we evaluate the model's performance to assess whether it is successful in adapting the granted resources to the task's and DAG's requirements. We share the obtained empirical results comparing \sscheduler\/ to the baseline schedulers. Then, we analyze the achieved performance gains by showing the schedule map for \sscheduler\/ pertaining to a specific task type and location, and how locality is preserved adaptively. Also, we explain the gains by relating them to the proportion of the cumulative work time to the overall scheduling time.

\begin{figure}[]
\centering
\subfigure[Memory-intensive task chain - Data within L1 cache. Initial NUMA node = 1. Initial thread = 31]{ \includegraphics[width=0.46\columnwidth]{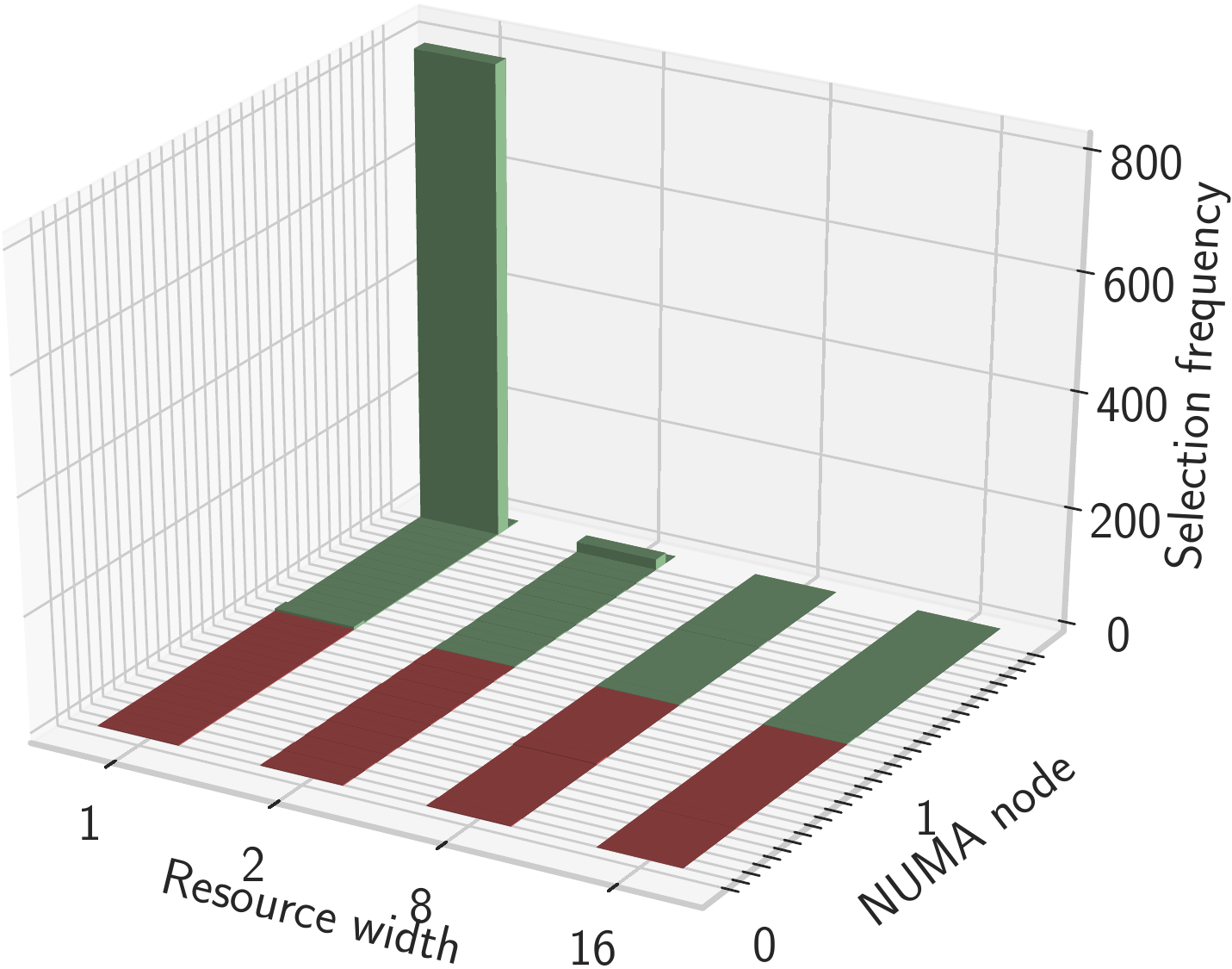}
\label{fig_l1_latency_model_acc}}\hspace{3pt} 
% \rulesep
\subfigure[Memory-intensive task chain. Initial NUMA node = 1. Initial thread = 17. Data fits in L3 cache]{ \includegraphics[width=0.46\columnwidth]{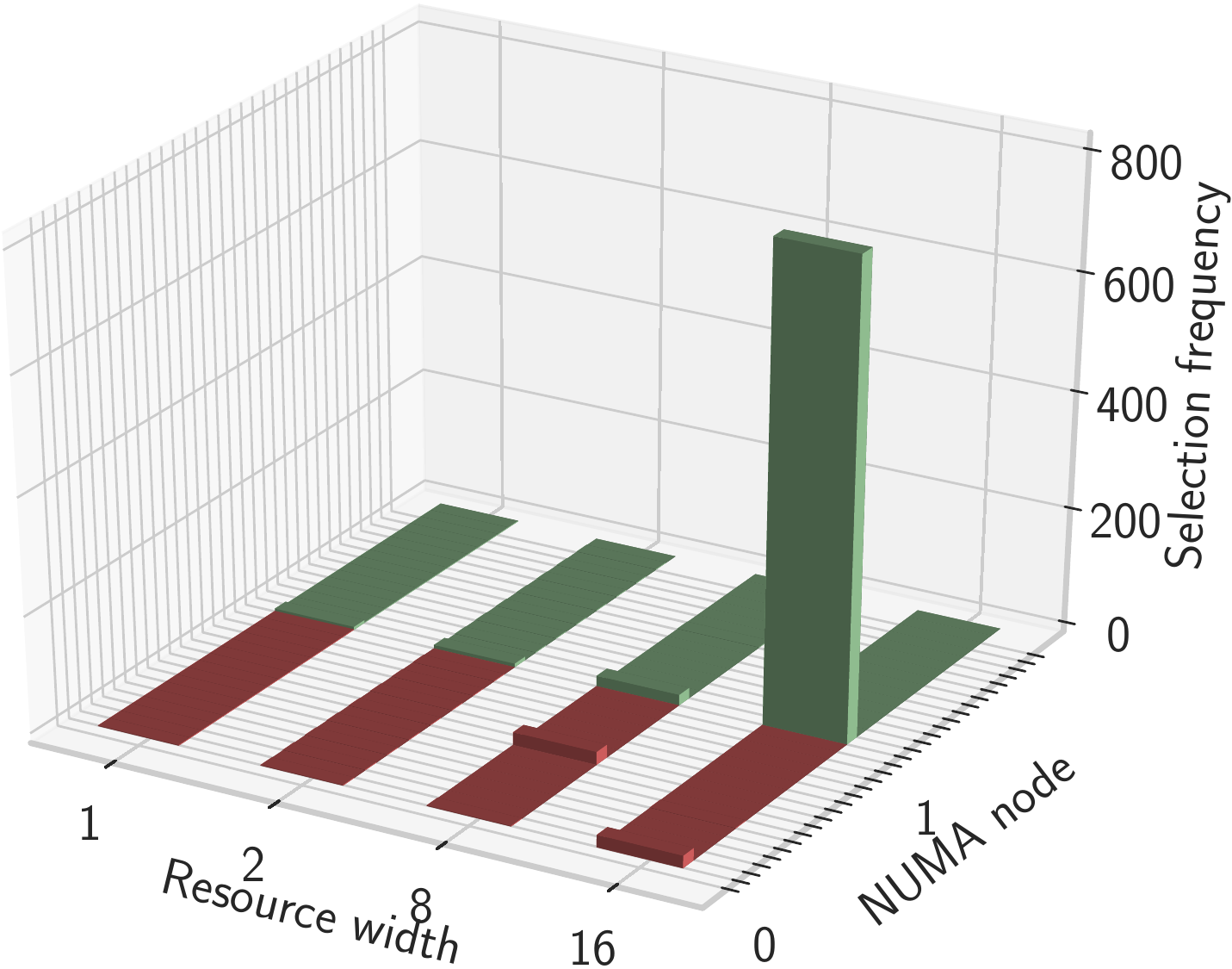}
\label{fig_socket_latency_model_acc}}\\
%\centering
\subfigure[Compute-intensive task chain. Initial node = 0. Initial thread = 0. Data fits within 2xL1 cache]{ \includegraphics[width=0.46\columnwidth]{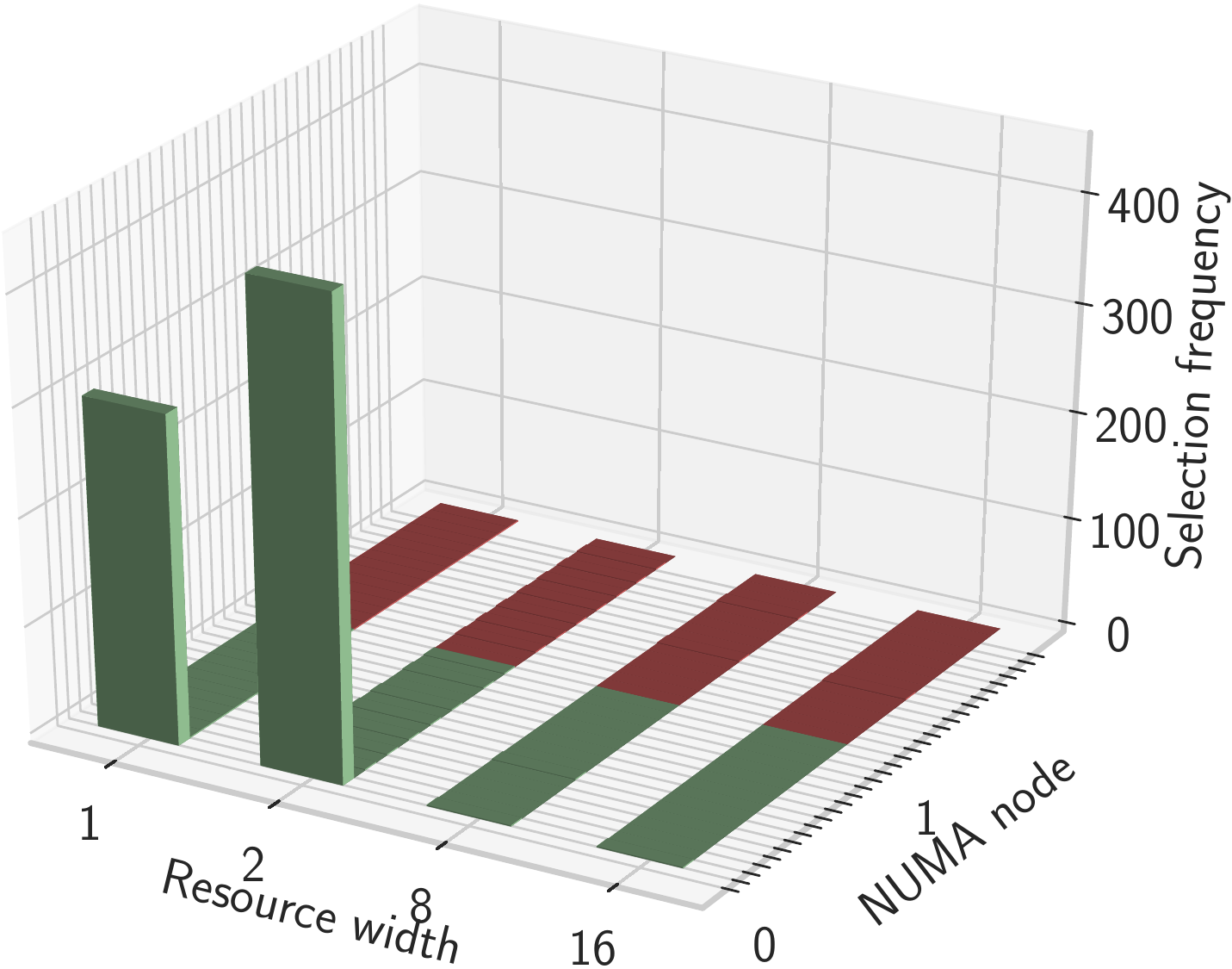}
\label{fig_compute_model_acc}}\hspace{3pt}  
% \rulesep
\subfigure[Compute-intensive task chain. {Initial node = 1. Initial thread = 17}. Data fits within L3 cache]{ \includegraphics[width=0.46\columnwidth]{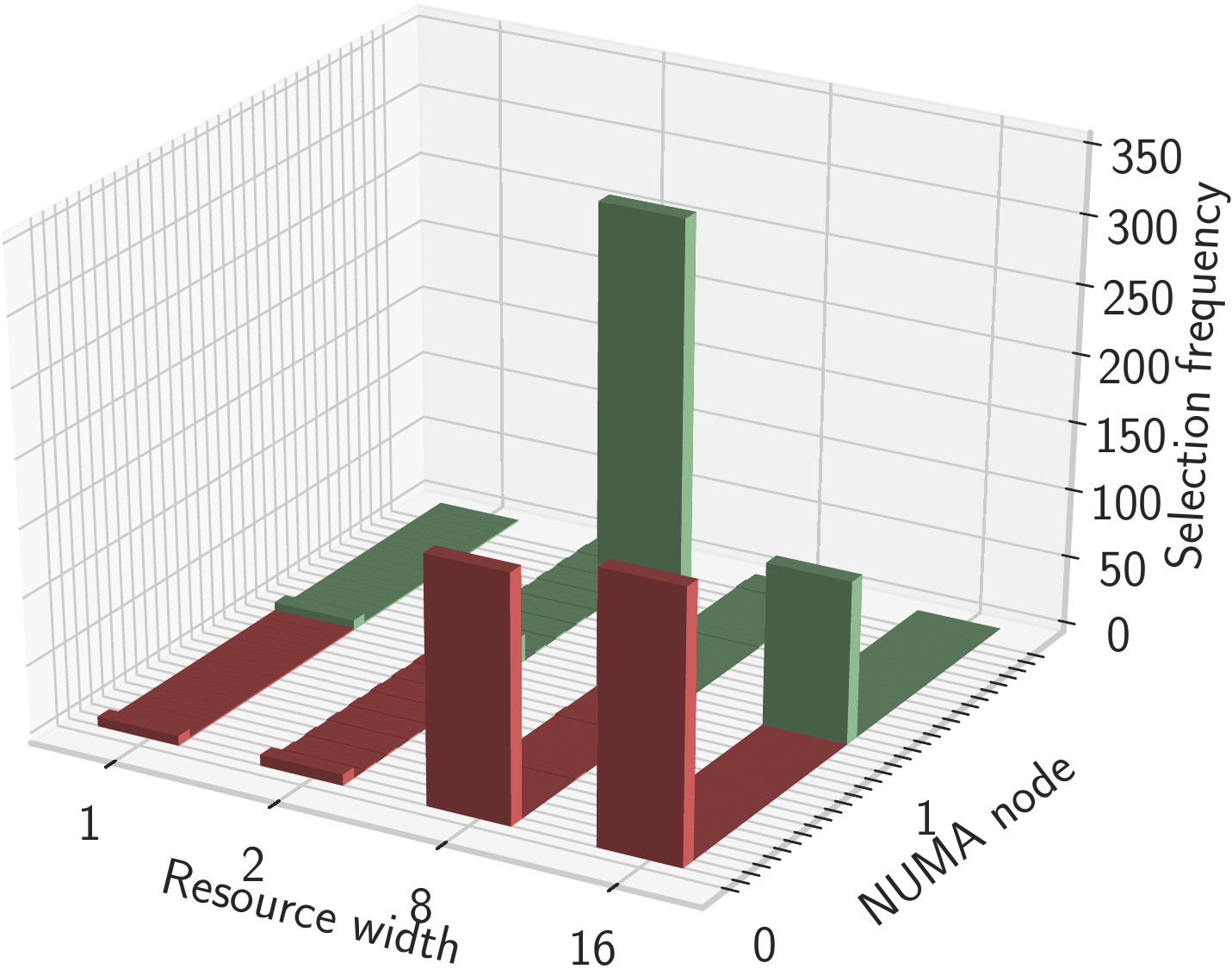}
\label{fig_large_compute_model_acc}} 
\caption{Resource selection model's behavior on \intelskl\/ for a chain of tasks .}
\label{fig:model_acc}
\end{figure}

\subsection{The Adaptive Resource-Moldable Selection}

We refer back to the highlighted motivational example discussed in Section~\ref{subsec:motivation} discussed in Figure~\ref{fig:nbody_size_vs_mflops_work_time}, where we see that locality maximizing using a single threaded task may result in a sub-optimal schedule. Not only \sscheduler\/ is able to place the dependencies on an efficient target NUMA node, but it is also able to maximize cache utilization by the online tuning of the resource width to match the task's requirement. Therefore, we evaluate whether the scheduler is able to adaptively tune locality depending on the initial location and size of the task's data.  
% The experiments shown in Figures~\ref{fig_l1_latency_model_acc} and ~\ref{fig_socket_latency_model_acc} map to the workloads indicated by Figures~\ref{fig:copy_small_abu_dhabi} and~\ref{fig:copy_large_abu_dhabi}, respectively. 

In this experiment, we create a chain of task dependencies exhibiting streaming behavior as shown in Figure~\ref{fig:synthbench}. The DAG is composed of as many chains as there are NUMA domains  (i.e. 2 chains in this experiment) and each performance model is referenced by its STA. As a simplification, each model reflects the tasks whose data belong to a distinct NUMA node. Each chain's head task and data are initially pinned to a NUMA domain. Figure~\ref{fig:model_acc} displays the schedule trace for a single chain of 1000 tasks . The $X$-axis is the resource width, the $Y$-axis is the thread, and the $Z$-axis is the frequency of selection. To simplify the $Y$-axis, we label the NUMA node id that encloses the set of 16 threads in the id range. Also, we color the node where the task is initialized in green.\\ 
\textbf{Memory-Intensive Schedule Map}: in Figure~\ref{fig_l1_latency_model_acc}, we expect that the dependent tasks will be scheduled to fit in the private L1 cache of the STA-mapped thread ($R=[LR=31, W=1]$). Also, we expect that since the tasks are latency-bound, the scheduler should decide most of the time to preserve the locality. Using Algorithm~\ref{alg:lagres}, this is evident in more than $90\%$ of the scheduling decisions. The second to highest frequency is observed for $R=[LR=30, W=2]$, which is still reasonable as the data size is exactly at the limit of the L1 cache. To test the automatic resource adaptation capability, we increase the problem size such that it does not fit in the private L2 cache (>=1024Kb) as shown Figure~\ref{fig_socket_latency_model_acc}. In this case, the scheduler opts for selecting the entire NUMA node (colored in green) to utilize the L3 cache totalling 22 MB ($R=[LR=16, W=16]$), while maximizing locality at the NUMA level.\\
% Figure~\ref{fig_scala_latency_model_acc} shows the performance of \lags\/ and \sscheduler\/ versus the baseline schedulers with tasks sized around 10MB ($>L3$) on \amdad\/. The $X$-axis is the number of tasks, whereas the $Y$-axis is the time to completion of the chain DAG application. The length of the chain is fixed, however, each increase in the number of tasks is done by doubling the number of chains from previous step. This results in an inherent load-imbalance (ready tasks >> number of threads). Utilizing the STA-based online model  shows a significant improvement over \laws\/ and \rwss\/ due to the adoption of the work-balancing scheme of the selection model algorithm. Additionally, \sscheduler\/  molds resources to avoid resource-level interference and results in around $5\%-10\%$ improvement, which is an acceptable percentage considering the uniformity of the DAG and the homogeneity of the underlying tasks.
\textbf{Compute-Intensive Schedule Map}: we also check how \sscheduler\/ adjusts the schedule up to the compute/resource requirements of a chain of compute-heavy tasks which constitute a single-precision direct $N$-Body code \cite{direct-n-body-Spur99}. In the small case shown by Figure~\ref{fig_compute_model_acc}, where tasks exceed the private L1 cache size (2xL1), the scheduler conservatively picks the locality maximizing places with width  ($W=1,2$). However, the chain of large tasks (Figure~\ref{fig_large_compute_model_acc}), is spread within (green region) and outside (red region) of the NUMA node containing the data, and the largest available resource widths ($W=8,16$) are chosen to maximize floating point capability. 
\subsection{Parallelism vs Locality under \sscheduler\/}

\begin{table}[t]
\caption{The resource width percentage choice as the DAG parallelism changes for a compute-intensive chain of tasks identified by an STA}
% \scalebox{0.75}{
\begin{tabular}{@{}|c|cccccccc@{}}
\toprule
\multicolumn{9}{|c|}{\textbf{The DAG Parallelism}} \\ \toprule
\textbf{Wid} & \multicolumn{1}{c|}{\textit{2}} & \multicolumn{1}{c|}{\textit{4}} & \multicolumn{1}{c|}{\textit{8}} & \multicolumn{1}{c|}{\textit{16}} & \multicolumn{1}{c|}{\textit{32}} & \multicolumn{1}{c|}{\textit{64}} & \multicolumn{1}{c|}{\textit{128}} & \multicolumn{1}{c|}{\textit{256}} \\ \midrule
\textit{1} & 0.1 & 0.3 & 0.5 & 1 & 1.9 & \cellcolor[HTML]{EFEFEF}\textbf{65.8} & \cellcolor[HTML]{C0C0C0}\textbf{96.1} & \cellcolor[HTML]{C0C0C0}\textbf{95.4} \\ \cmidrule(r){1-1}
\textit{2} & 0.1 & 0.2 & 0.3 & \cellcolor[HTML]{C0C0C0}\textbf{97.8} & \cellcolor[HTML]{C0C0C0}\textbf{94.2} & \cellcolor[HTML]{EFEFEF}\textbf{29} & 1.3 & 0.7 \\ \cmidrule(r){1-1}
\textit{4} & 0.1 & \cellcolor[HTML]{C0C0C0}\textbf{99} & \cellcolor[HTML]{C0C0C0}\textbf{98.9} & 0.6 & 1.3 & 3.9 & 0.6 & 0.7 \\ \cmidrule(r){1-1}
\textit{8} & \cellcolor[HTML]{C0C0C0}\textbf{99.7} & 0.4 & 0.5 & 0.6 & 2.6 & 1.3 & 1.9 & 3.3 \\ \cmidrule(r){1-1}
\end{tabular}
% }
\label{tbl:dop_vs_width}
\end{table}
One of the important properties to study with respect to dynamic locality-aware scheduling is how the scheduler balances the trade-off between locality and parallelism, which is a known challenge in scheduling task-based programs. This is because of the fact that increasing the number of threads decreases the apparent spatial locality since access streams from independent cores are interleaved~\cite{locality_parallelism_Jeong12}. Since \sscheduler\/ addresses this trade-off by adjusting the task's resource width, we analyze how it reacts to a wide range of DAG parallelisms spanning (2 - 256). The change in the DAG parallelism is highly evident in divide-and-conquer computations (a special case of fully-strict computations \cite{blumofe-jacm99}) as we go up the recursion tree, where parallelism gets lower. Also, the DAG parallelism changes due to factors such as load-imbalance, dependency checking, synchronization and so on.\\
\textbf{Impact of Changing the DAG Parallelism}: to simplify the analysis, we see how \sscheduler\/ behaves with different DAG parallelisms using the synthetic benchmark depicted by Figure~\ref{fig:synthbench}. We fix the number of tasks to 50k, so that we are able to monitor the changes in performance as a function of the DAG parallelism as shown in Figure~\ref{fig:chain_locality_parallelism}, with N=128 for the MatMul case and N=512 for the Stream Copy case. 
Then, we compare to \adws\/ as a representative of locality-aware schemes. For the challenging case of lower parallelisms (2 - 8), \sscheduler\/ outperforms \adws\/ by approximately (3.5$\times$, 3$\times$, 2.5$\times$) in the compute and memory-intensive cases shown in
Figures~\ref{fig:chain_compute_locality_parallelism} and ~\ref{fig:chain_memory_locality_parallelism}, respectively. However, the gap is closed as the machine hardware parallelism is reached, with a slight degradation for higher parallel slackness than 32. This shows that \sscheduler\/ has a sustainable behavior across different parallelisms and is able to effectively adapt to the case when locality preserving is not enough to achieve higher throughput. Since scheduling decisions are made on a per-task basis following the depicted online performance model, the aggregate effect on performance of mixing Copy and MatMul tasks combines the trends from the individual cases as shown by Figure~\ref{fig:chain_mixed_locality_parallelism}. \\
\textbf{Analyzing \sscheduler's Performance Gain}: to demonstrate the sources of the gain, Table~\ref{tbl:dop_vs_width} prints the trace for the resource width choice of a single chain of MatMul task (N=128) across different runs that differ by the DAG parallelism. Since \sscheduler\/ minimizes the parallel cost ($T(P) \times w$), it is able to dynamically aggregate resources (i.e. increase the width) to the task at a lower DAG parallelism. For example, \sscheduler\/ detects that the parallel cost of using 8 threads is lower when the DAG parallelism is 2 in 99.7\% of the cases. Note that the leader thread of this task has an id of 8 as depicted by the STA. A step-wise increase in the task resource width occurs until the DAG parallelism is 32 which matches the machine's parallelism. Beyond this point, the automatic width choice is 1. 
In the following sections, we study and analyze the effectiveness of these techniques in the context of different classes of applications.
% Since the proposal is resource moldable, we expect it fulfill the gap in lower parallelism by increasing the width.
\subsection{Application Performance Evaluation}
\begin{figure}[t]
\centering
\subfigure[2D Stencil - Time]{ \includegraphics[width=0.47\columnwidth]{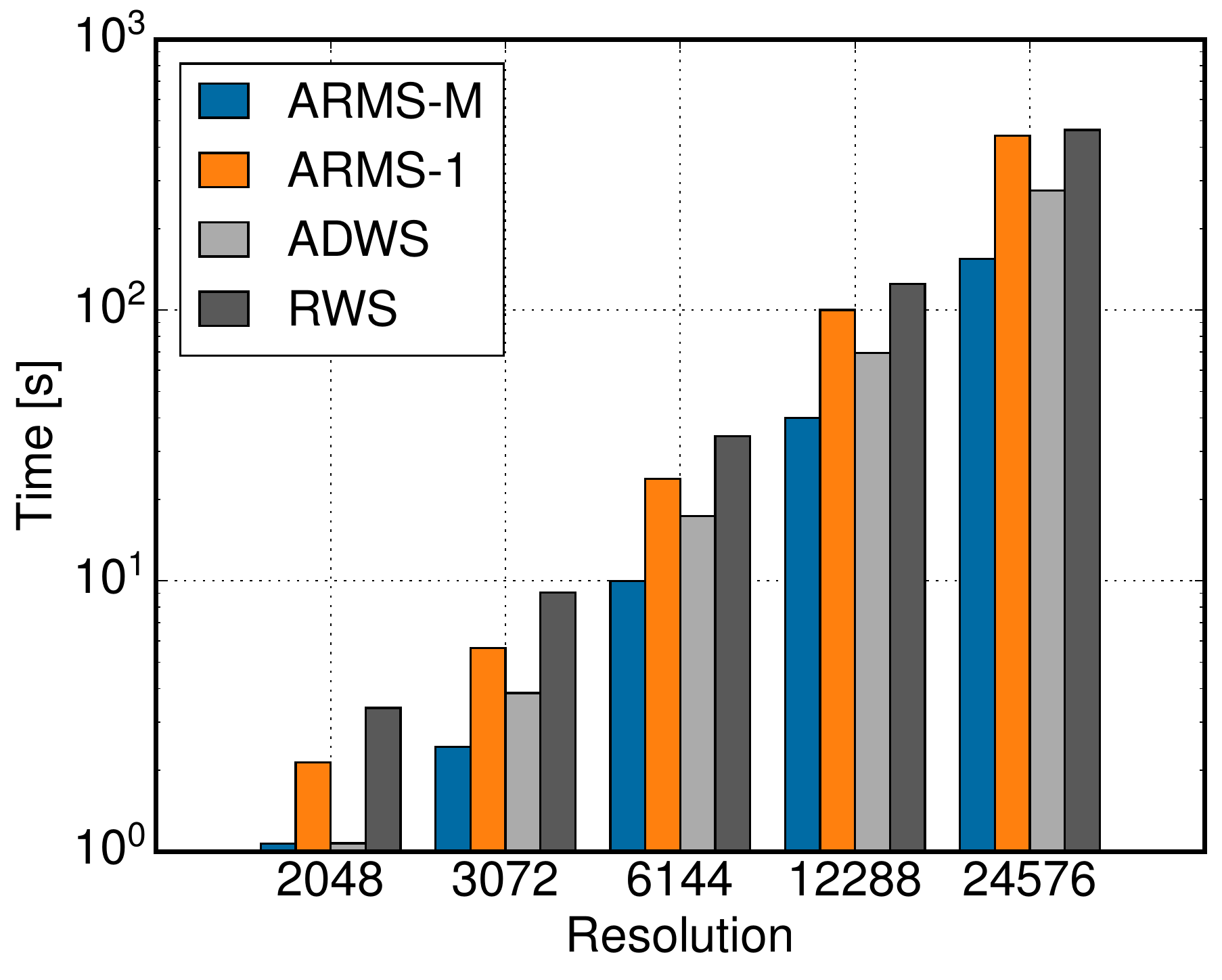}
\label{fig:heat_barplot}} 
\subfigure[Recursive MatMul - Time]{ \includegraphics[width=0.47\columnwidth]{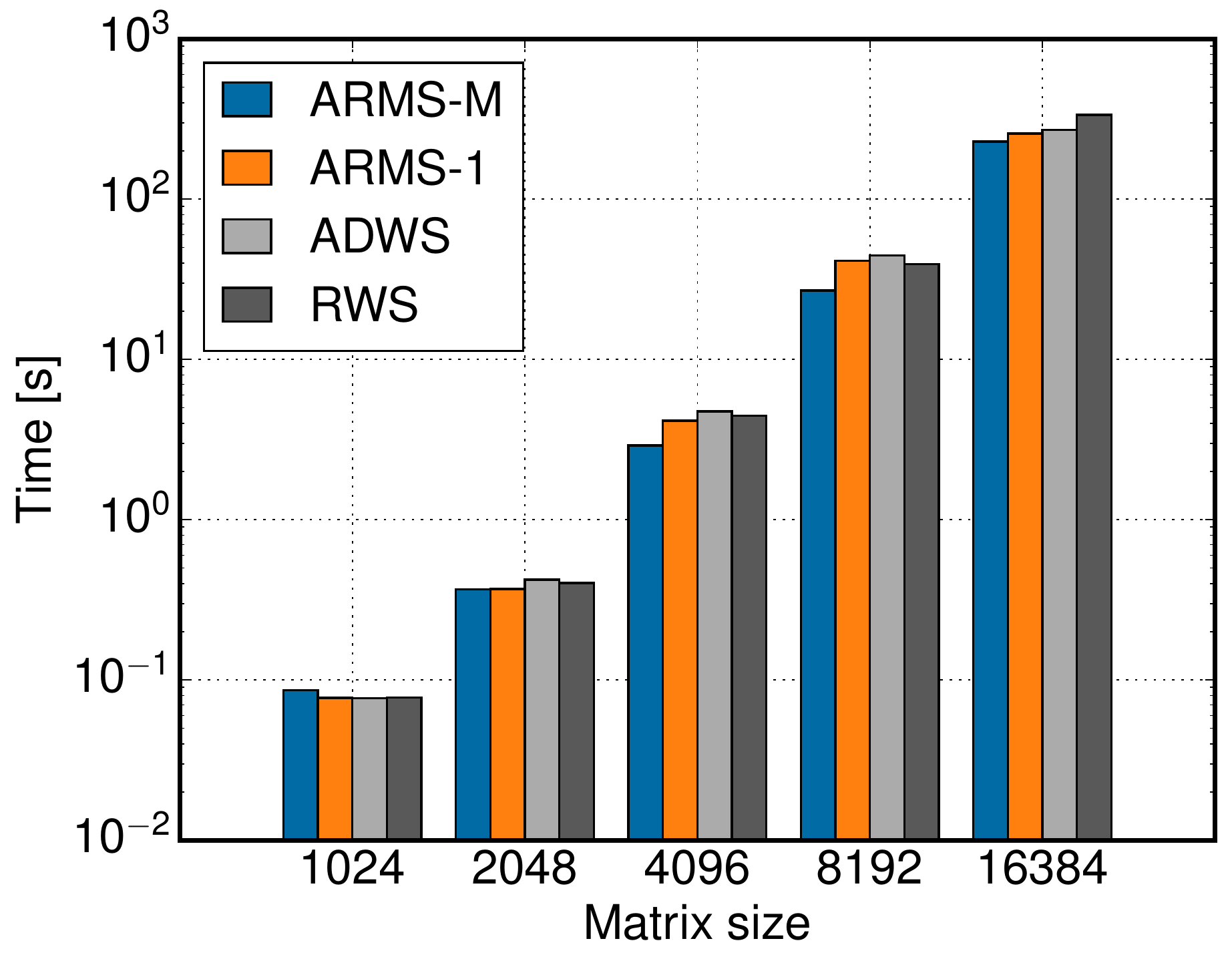}
\label{fig:matmul_barplot}}\\
\subfigure[FMM - Time]{ \includegraphics[width=0.47\columnwidth]{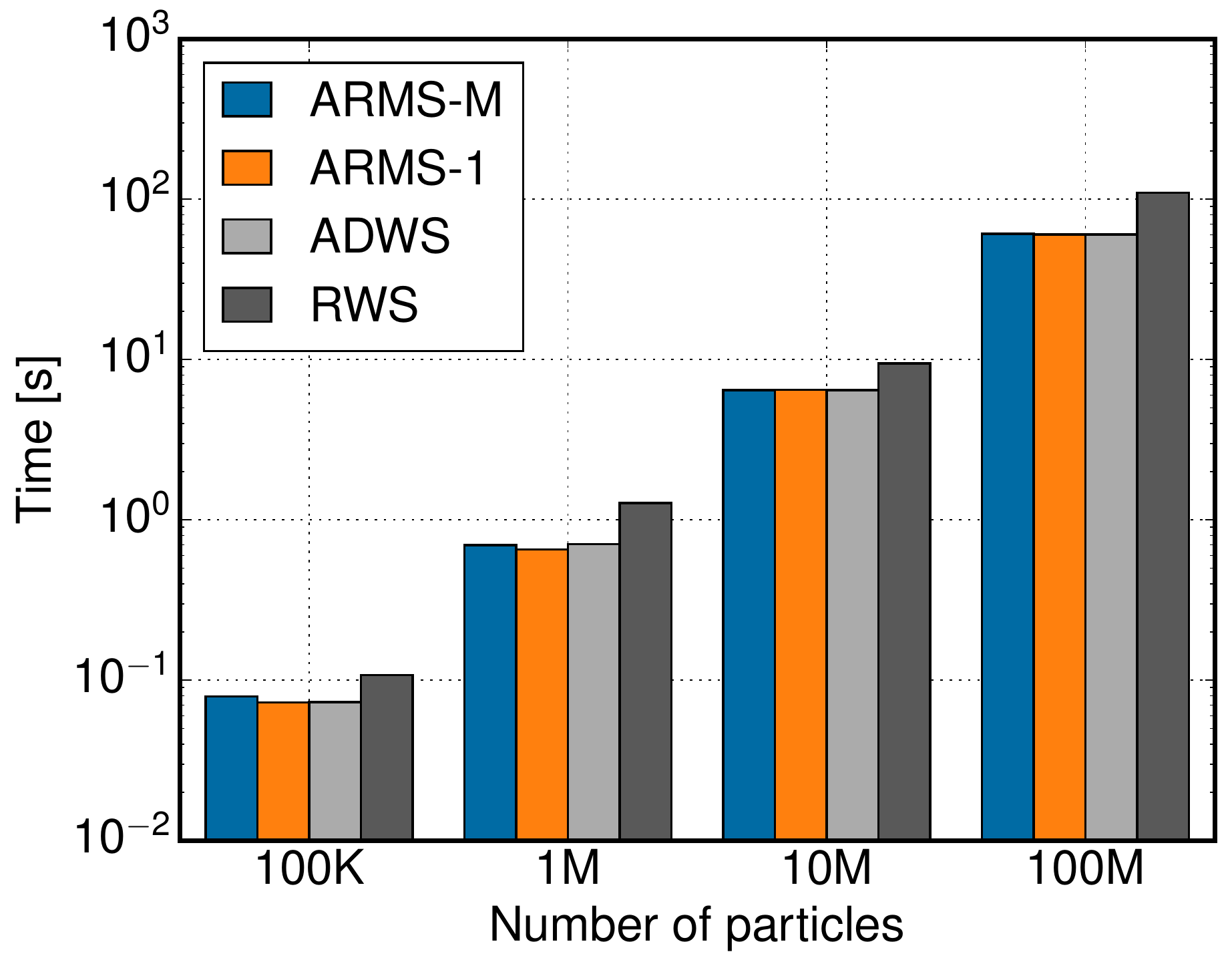}
\label{fig:fmm_barplot}} 
\subfigure[SparseLU - Time]{ \includegraphics[width=0.47\columnwidth]{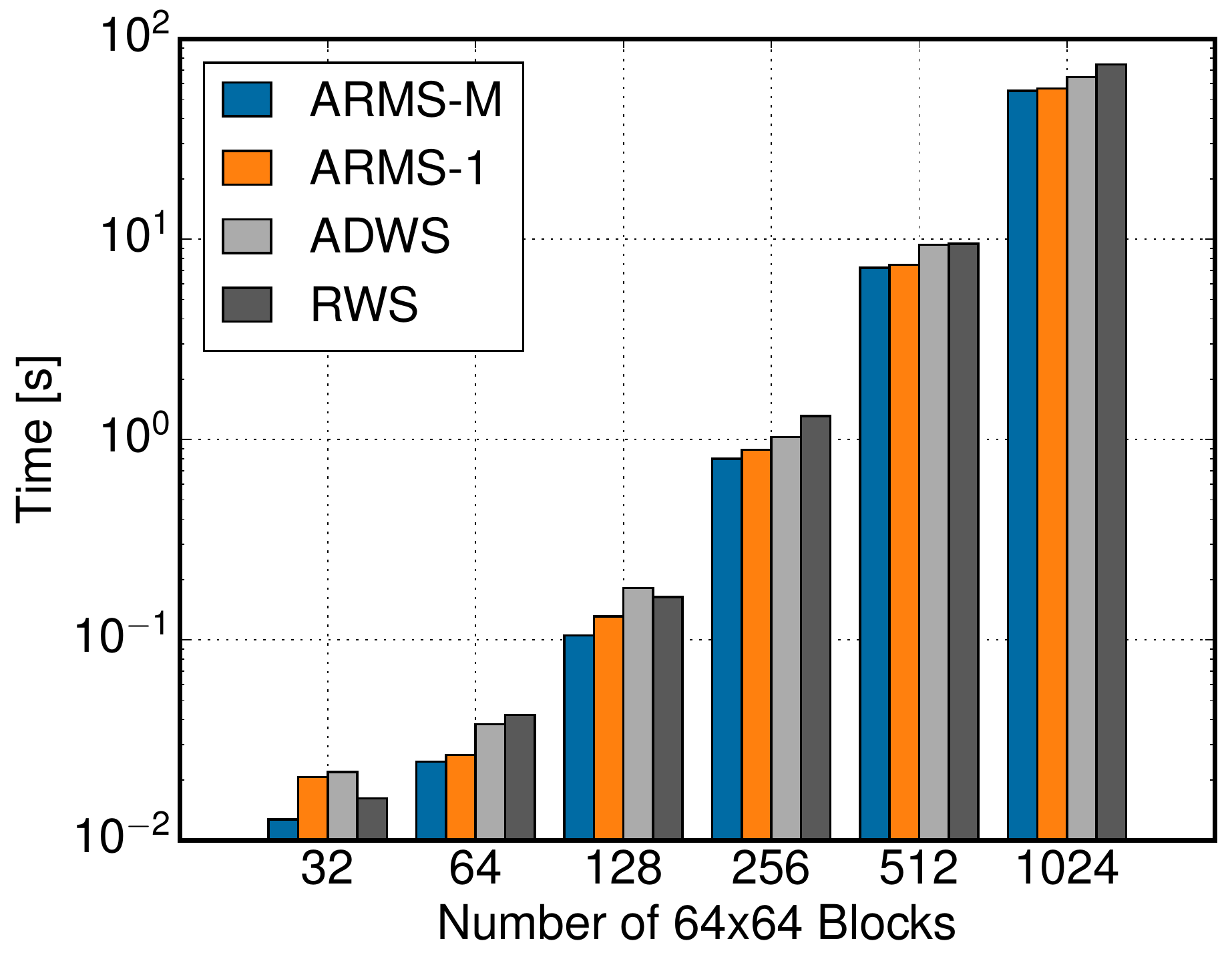}
\label{fig:sparselu_barplot}}
\caption{Performance comparisons for task-based applications}
\label{fig:par_time}
\end{figure}

\begin{figure}[t]
\subfigure[2D Stencil - L2 misses]{
\includegraphics[width=0.47\columnwidth]{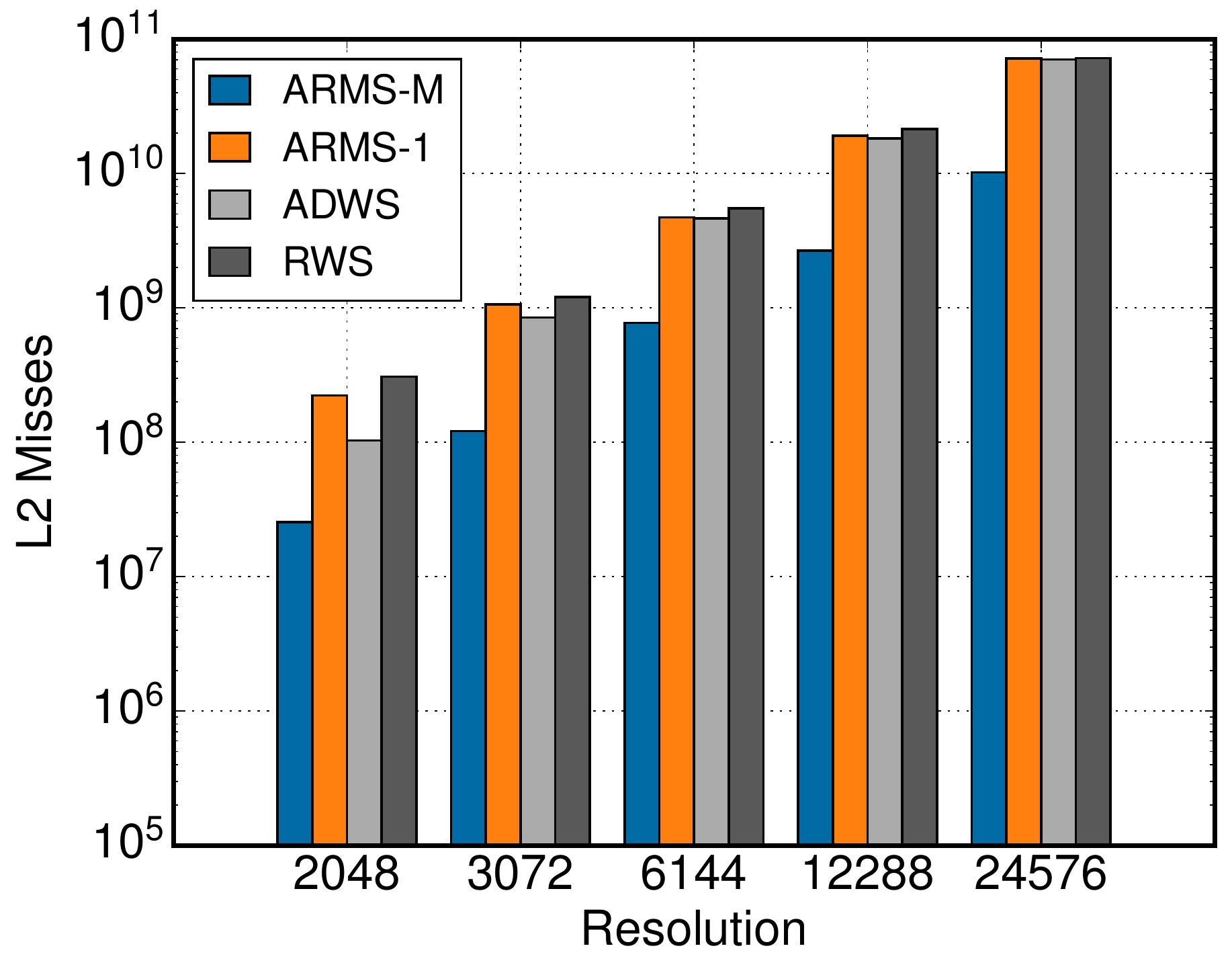}
\label{fig:heat_barplot_l2misses}
}
\subfigure[Recursive MatMul - L2 misses]{
\includegraphics[width=0.47\columnwidth]{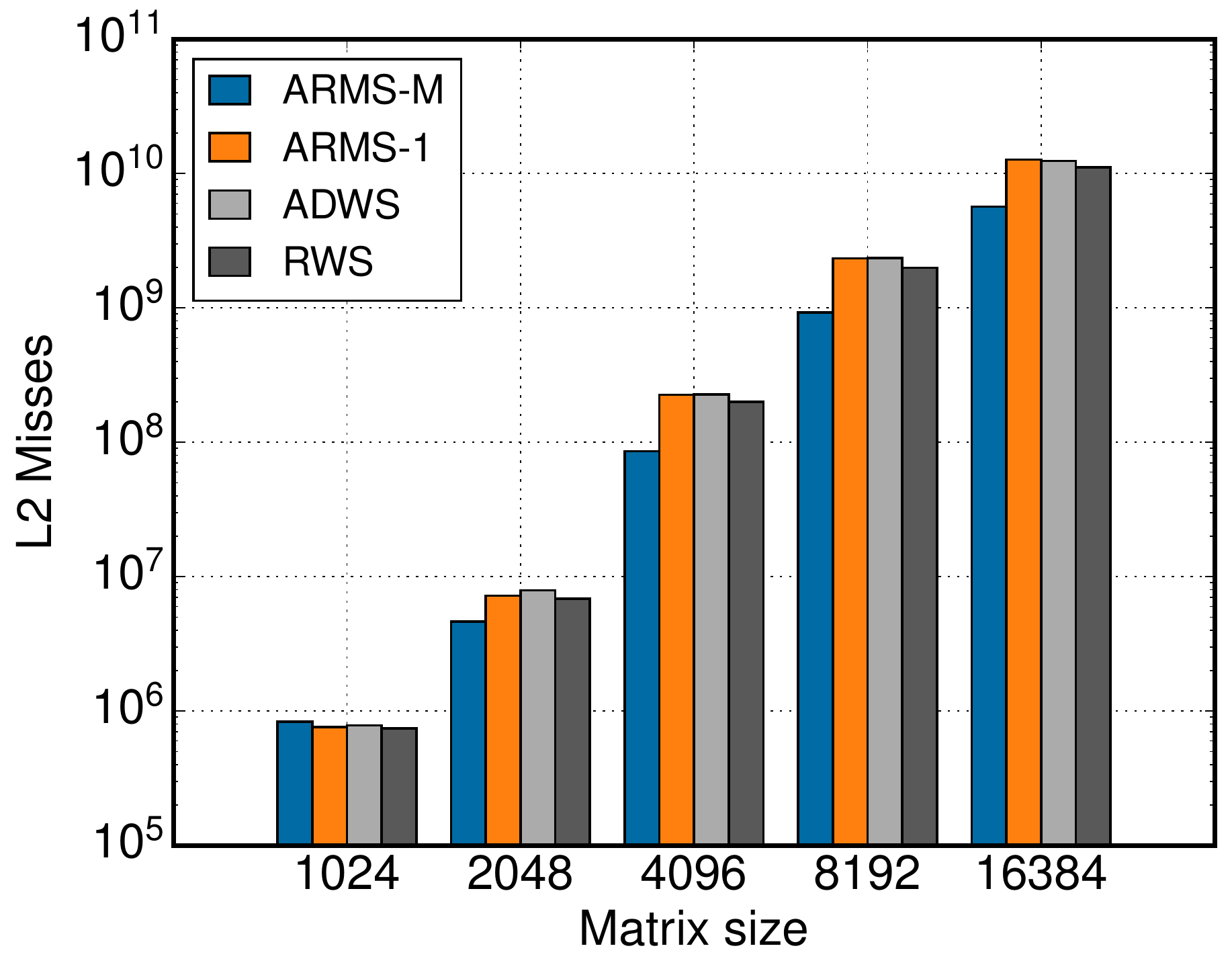}
\label{fig:matmul_barplot_l2misses}
}
\caption{L2 misses mapped to the performance gains that \sscheduler\/ achieves in the 2D Stencil and Recursive MatMul applications. }
\label{fig:heat_matmul_l2_misses}
\end{figure}

In this section, we evaluate the performance of \sscheduler\/ against the baselines schedulers using various application DAGs. The applications and baselines have been described in Section~\ref{sec:methodology}. We also assess the impact of
moldability on \sscheduler\/ by studying the two scheduling variants : \armsn\/ and \lags\/. 
This evaluation intends to showcase the consistent enhancement that \sscheduler\/ exhibits across different application classes. The granularity of the task creation across all benchmarks have been configured to be in the range of (2 - 4) L1 caches (64Kb - 256Kb) and within a private L2 cache ($<$ 2048Kb). The initial resource width is set to 1 for all tasks (1-to-1 assignment). \\
\textbf{2D-Stencil}: this DAG consists of a copy task and a 5-point stencil compute grid task (Figure~\ref{fig_dag_heat}) iterating 2000 times. Experiments incrementally double the mesh resolution. This application has a clear data reuse pattern across the iterations, however, the data does not fit within a single L1 cache, so moldability has a consistent improvement from 1.5x - 2x over the best baseline (\adws\/) as shown by Figure~\ref{fig:heat_barplot}. Based on our traces, \sscheduler\/ molds to 2 cores in more than 90\% of the scheduling decisions of the stencil compute tasks. The improvement also maps to up to an order of magnitude reduction in the application's L2 misses (Figure~\ref{fig:heat_barplot_l2misses}). \\
\textbf{MatMul and SparseLU}: in the case of MatMul, the improvement takes place at a relatively larger matrix size. Since we set the leaf block size to 128, the first two cases result in 3 - 4 subdivisions, which is not enough to train the model. However, \sscheduler\/ still {performs better than} \rwss\/ and \adws\/ from a matrix size of 2048 (Figures~\ref{fig:matmul_barplot} and ~\ref{fig:matmul_barplot_l2misses}). Similar improvement, shown in Figure~\ref{fig:sparselu_barplot}, are achieved in SparseLU with 64x64 blocks. \\
\textbf{FMM}: Last but not least, considering a highly irregular DAG structure like the FMM's DAG (Figure~\ref{fig_fmm}), \armsn\/, \lags\/ and \adws\/ behave like locality-aware work-stealing schedulers due to the high parallelism and compute intensity of FMM kernels as we can observe from Figure~\ref{fig:fmm_barplot}. However, this only shows that \sscheduler\/ does best effort at either matching or outperforming the baselines without user-level computational hints or prior workload assumptions.

{\sscheduler\/  is effective in providing better or comparable schedules to the state-of-the-art. Molding a task to multiple threads in a locality aware-manner is especially profitable in applications with uniform data reuse patterns such as Stencil and recursive MatMul (N>=2048) as demonstrated from the behaviors in Figure~\ref{fig:heat_matmul_l2_misses}. However, due to constructing locality-adaptive performance model, \armsn\/ behaves like non-moldable locality-aware schedulers (\lags and \adws\/) in FMM and SparseLU. \armsn\/ automatically achieves this without needing to change the user-code or having to change the scheduling options. 
}
% \textbf{Analysis of \ar}: 
% In the case of 2D stencil (Figure~\ref{fig:heat_barplot}), we achieve 

% \begin{figure*}[t]
% \centering
% \subfigure[Stencil - \intelhw]{ \includegraphics[width=0.3\textwidth]{figs/experiments/Total_CPU_time_hebbe.pdf}
% \label{fig_total_cpu_time_hebbe}} 
% \rulesep
% \subfigure[Stencil (Compute) - \amdad - tasks data pinned to Node 6 - leader thread = 42]{ \includegraphics[width=0.3\textwidth]{figs/experiments/copy_scatter_puck.pdf}
% \label{fig_stencil_compute_numa_auto}} 
% \rulesep
% \subfigure[Stencil (Copy) - \intelhw - tasks data pinned to Node 0 - leader thread = 0]{ \includegraphics[width=0.3\textwidth]{figs/experiments/copy_scatter_hebbe.pdf}
% \label{fig_stencil_copy_numa_auto}} 
% \caption{Comparison of the cumulative work/other distribution, and the scheduling trace for a Stencil (compute and copy) tasks. The denoted tasks have have a single software topology address. Mesh resolution = $2048\times2048$.}
% \label{fig:stencil_work_other_time}
% \end{figure*}

% \subsection{Analysis}

% \begin{figure*}[!t]
% \centering

% % \subfigure[Stencil - \amdad]{ \includegraphics[scale=0.35]{figs/experiments/Total_CPU_time_puck.pdf}
% % \label{fig_total_cpu_time_puck}} 
% \caption{The cumulative work/other distribution executed by each thread using the traditional versus the proposed approach.}
% \label{fig:stencil_work_other_time}
% \end{figure*}

% \input{literature}
\section{Related Work}
\label{sec:related_work}
There has been extensive literature addressing data locality concerns in HPC applications and systems. On the theoretical level, models that predict locality for parallel programs by means of reuse distance analysis have been discussed in ~\cite{ding-pldi03, ding-mrtr09,cascaval-ics03}. Such models aid the programmers in understanding the memory access patterns of their codes. {Additionally, data-centric programming models (e.g.  Legion~\cite{legion-bauer-sc-12}) and language extensions (e.g. Hierarchical Place Trees (HPT)~\cite{hpt-yonghong-lcpc-2010}) require the domain expert programmer to analyze the data access regions and express the program using their interfaces, and to make explicit assignments to execution places.}
Despite their effectiveness, programmer level techniques (e.g. loop tiling or hierarchical cache optimizations~\cite{Kowarschik2003}) are known to have limited portability across platforms. Hence, high-level abstractions in data locality have received a growing attention from the HPC research community~\cite{unat-tpds17,bauer-sc12}. These abstractions are meant to adopt existing features from parallel execution frameworks, such as the affinity partitioner and task arenas from Intel TBB (Thread Building Blocks), places and processor binding from OpenMP, topology and communication reducing in MPI (Message Passing Interface)~\cite{bosilca-ipdpsw11, abduljabbar_isc17_comm_reducing_hsdx}.

As discussed, many task-based parallel libraries embrace random work-stealing scheduling mechanisms, which are provably known to be effective in maximizing dynamic load-balancing. However, they are bound to suffer from scalability issues with memory-bound tasks. To address these, techniques have been proposed to achieve locality-aware work-stealing behavior such as the Almost Deterministic Work Stealing (\adws\/)~\cite{acct-19-vikranth}. This is done via deterministic mapping of tasks to resources that is applied based on the programmer's annotation of workload size. Also, the paper discusses an approach to achieve hierarchical localized work stealing. Another technique is highlighted in Scalable Locality-aware Adaptive Work-stealing Scheduler (SLAW), which also allocates the resources based on user annotations of locality (a place index)~\cite{guo-ipdps10}. Last but not least, XKaapi~\cite{gautier-ipdps13} is a locality-aware work stealing scheduler that supports heterogeneous data flow programming targeting both CPUs and GPUs. 

While these are effective techniques in reducing the side effects of random work-stealing, none of them consider the difficulty of obtaining a static classification of applications into memory and compute intensive classes. We think that \sscheduler\/ is not completely orthogonal to the literature above, as it still maximizes locality when deemed necessary by the online model.  However, \sscheduler\/ is not yet-another locality-aware scheduler, but it rather adopts a dynamic strategy and opts for locality maximizing by understanding the behavior of a task at runtime. We believe that this strategy is only more realistic considering the evolving complexity of applications and hardware.

\section{Conclusions}
\label{sec:conclusion}
In this paper, we presented a novel scheduling strategy that dynamically tunes its locality-awareness. This is achieved without statically establishing locality-aware decisions irrespective of the underlying task's arithmetic and memory footprints, a choice that can lead to suboptimal performance. 
By adopting an online performance model paired with platform-independent topology information,
%By adopting an online model per STA, 
the scheduler is able to maximize locality for the latency-intensive tasks, and to relax it for the compute-intensive counterpart. The resource moldability feature allows to adapt the granted resources to the task's requirements and to reduce the cache misses, and achieves performance gains %based on our findings using 
on a variety of application DAGs.
% {The gains are achieved without prior knowledge of the underlying hardware or explicit pinning of tasks to hardware places to maintain locality.}
% So at the end, \sscheduler\/ comes to the rescue, and ``task'' accomplished!.
%In the cases of lower DAG parallelism, \sscheduler\/ has been effective in addressing the trade-off between locality and parallelism by increasing the task resource width. This resulted in a clear improvement compared to locality-aware baseline (\adws\/).
% Baseline schemes such as \rwss\/ and \laws\/ have been shown ineffective when tasks have larger working sets. Such scenarios result in higher Cache/NUMA miss penalties and resource over-subscription.

% More work is needed to reduce the scheduling overhead for fine-grain tasks, which are essential to boost concurrency. Also, generalization of  the notion of STA is still required to cover applications where topology is not directly present or data is not spatially blocked. For example, it is not trivial to use \sscheduler\/ when the DAG operates on a single large array of data, and the inner tasks just maintain indices to the positions.  Nonetheless, the model is designed to be extensible and can accommodate locality driven techniques without a considerable effort. 
% The code and the presented experiments will be made available on \texttt{GitHub}.

\bibliographystyle{ACM-Reference-Format}
%%\bibliographystyle{IEEEtran}
%%\bibliography{ref}
\bibliography{jrlib-miquel,ref}

\end{document}